\documentclass[prd,aps,amsmath,amssymb,nofootinbib]{revtex4}

\usepackage[english]{babel}
\usepackage[utf8]{inputenc}
\usepackage{graphicx,graphics}
\usepackage{dcolumn}
\usepackage{bm}
\usepackage{hyperref}

\usepackage{comment}
\usepackage{blindtext}
\usepackage{braket} 
\usepackage{indentfirst}
\usepackage{setspace}
\usepackage{caption}
\usepackage{subcaption}
\DeclareMathOperator{\tr}{tr}

\usepackage{mathrsfs}
\usepackage{pstricks}
\usepackage{color}
\usepackage{slashed}
\usepackage{amsmath}
\usepackage{epsfig}
\usepackage{amsfonts}
\usepackage{amssymb}

\def\beq{\begin{equation}}
\def\eeq{\end{equation}}
\def\bea{\begin{eqnarray}}
\def\eea{\end{eqnarray}}
\def\nn{\nonumber}

\def\Tr{\textrm{Tr}}

\usepackage{hyperref}
\usepackage{xcolor}
\hypersetup{
    colorlinks,
    linkcolor={red!80!black},
    citecolor={blue!50!black},
    urlcolor={blue!80!black}
}

\begin{document}

\title{Reduced quantum electrodynamics in curved space}
\author{P. I. C. Caneda}
\email{caneda@cbpf.br}
\affiliation{Centro Brasileiro de Pesquisas F\'isicas, 222990-180 Rio de Janeiro, RJ, Brazil}
\author{G. Menezes}
\email{gabrielmenezes@ufrrj.br}
\affiliation{Centro Brasileiro de Pesquisas F\'isicas, 222990-180 Rio de Janeiro, RJ, Brazil}
\affiliation{Departamento de F\'isica, Universidade Fe\-de\-ral Rural do Rio de Janeiro, 23897-000 Serop\'edica, RJ, Brazil.}
%

\begin{abstract}

An approach that has been given promising results concerning investigations on the physics of graphene is the so-called reduced quantum electrodynamics. In this work we consider the natural generalization of this formalism to curved spaces. We employ the local momentum space representation. We discuss the validity of the Ward identity and study one-loop diagrams in detail. We show that the one-loop beta function is zero. As an application, we calculate the one-loop optical conductivity of graphene by taking into account curvature effects which can be incorporated locally. In addition, we demonstrate how such effects may contribute to the conductivity. Furthermore, and quite unexpectedly, our calculations unveil the emergence of a curvature-induced effective chemical potential contribution in the optical conductivity.
\end{abstract}

\maketitle

\section{Introduction}

In the last decades condensed-matter systems of diverse natures have been increasingly studied under the methods of quantum field theory (QFT). This has emerged as an important tool in the condensed-matter community in the sense that QFT allows us to theoretically explore the prominent physics developing on the relevant low-energy scale probed in experiments. Remarkably, it has also been realized that elusive particles that appear in the context of high-energy physics, such as Weyl and Majorana fermions, can naturally emerge in the form of quasi-particles in a condensed-matter setting~\cite{Perge:2014,Xu:2015}. On the other hand, recently one has witnessed the outbreak of investigations dedicated to collectively understand the prospects of exploiting condensed-matter models as possible experimental realizations of physical situations that arise in the context of general relativity and of quantum field theories in curved backgrounds. For instance, it now has been well established that kinematic aspects of black holes can be investigated in weakly interacting Bose gases~\cite{BHoles}. In this analog model configuration, theoretical surveys have also probed aspects of interesting kinematical effects that arise in classical and quantum systems, such as, for example, phenomena involving superradiance processes~\cite{SR}.

The investigation proposed here considers this current trend to borrow concepts originally developed in high-energy physics for the study of low-energy systems commonly found in condensed matter. We are particularly interested in the transport properties of graphene. The low-energy physics of two-dimensional carbon systems~\cite{1,2} is governed by the presence of two generations of massless Dirac fermions. The electronic interactions in Dirac liquids lead to a wealth of intriguing transport phenomena which have attracted a fair amount of attention since the first synthesization of graphene in $2004$~\cite{3}. Indeed, recent experiments uncover the relevance of such electronic interactions at low temperatures~\cite{5,6,7,8}. In turn, the interplay between strong Coulomb interactions and weak quenched disorder in graphene has also been elucidated, and the general expectation is that vector-potential disorder may play a key role in the description of transport in suspended graphene films~\cite{Foster:2008}. Motivated by clear evidence of the strongly coupled nature of graphene, transport coefficients were calculated within a modern holographic setup~\cite{Sachdev:2017}.  

The specific structure of the $2D$ crystal lattice permits graphene systems to be viable settings to study some of the interesting effects which arise in QFT in curved space-times~\cite{iorio:2012,iorio:2014,Gibbons:2012}. In this context, measurable effects of QFT in a curved-background description of the electronic properties of graphene represent a growing ongoing line of research. A number of proposals to interpret several observed effects in graphene sheets such as curved ripples~\cite{Vozmediano:2007}, corrugations~\cite{Vozmediano:2006}, pure strain configurations~\cite{iorio:2015} and even nonuniform elastic deformations~\cite{Arias:2015} in the light of a curved-space description of the electronic properties of graphene has occupied much of the contemporary associated literature. The appearance of gauge fields in graphene systems has also made it possible to establish a firm bridge between the physics of graphene and gravity-like phenomena allowing the unification of concepts from elasticity and cosmology~\cite{Vozmediano:2010}.

The chiral nature of the charge carriers in graphene is responsible for the existence of a minimal AC conductivity in the collisionless regime which is universal~\cite{Andrei:2008}. In this respect much theoretical effort has been devoted to understand the effects of electronic interactions on the optical conductivity in such a scenario (for an interesting discussion, see Ref.~\cite{teber} and references cited therein). One possible framework with which one can address this issue is given by the so-called reduced quantum electrodynamics (RQED). This is a quantum field theory describing the interaction of an Abelian $U(1)$ gauge field with a fermion field living in flat space-times with different dimensions~\cite{marino,Miransky:2001}. Motivations for the investigation of such reduced theories comprise their feasible application in low-dimensional condensed-matter settings, in particular graphene systems. Indeed, it has been claimed that calculations within the formalism of RQED reproduce as close as possible the experimental results for the minimum conductivity of graphene ~\cite{marino2}. Electromagnetic current correlation has also been computed within the context of RQED~\cite{Teber:2012}. Other interesting, noteworthy features of RQED include the validity of the Coleman-Hill theorem and the existence of quantum scale invariance~\cite{Dudal18, Dudal19}. For recent studies of chiral symmetry breaking in RQED at finite temperature and in the presence of a Chern-Simons term, see Refs.~\cite{Cuevas20,Olivares20}.

In the present exploration our theoretical laboratory will be the generalization of the formalism of RQED to curved spaces. We do not wish to single out one particular metric in our exploration, but instead we will keep our discussion to general spatial geometries. For that we will use a momentum-space representation of the Feynman propagator in arbitrary curved space-times~\cite{Birrell,Parker}. As usual the construction rests upon the usage of Riemann normal coordinates~\cite{Petrov,Parker:1979}. As an application, our discussion will allow us to calculate the one-loop high-frequency behavior of the optical conductivity in the presence of curvature effects in graphene by using the Kubo formula. We will demonstrate that the effect of the curvature upon the optical conductivity, to a certain extent, is to induce the appearance of an effective chemical potential when the Ricci scalar is positive. We will also explore the intriguing possibility that such curvature effects can actually contribute to an increase in the conductivity of graphene. We employ units such that $\hbar = c = 1$.

\section{RQED in curved space}

\subsection{RQED in flat space}\label{flatrqed}

Let us begin our discussion in flat space. Massless Dirac electrons are assumed to interact via the RQED in two spatial dimensions. Such a model in flat space is given by the following action\footnote{To avoid cluttering notation the space-time indices of both dimensions are labeled equally, their range is left implicit from their corresponding action.} (for the Euclidean version, see Ref.~\cite{marino2}):
\beq
S = \int d^{d_{\gamma}}x \left[ -\frac{1}{4}\,F_{\mu\nu} F^{\mu\nu} 
- \frac{1}{2\xi} (\partial_{\mu} A^{\mu})^2 \right] 
+ \int d^{d_{e}}x \Bigl( \bar{\psi}_{A}\,iv_{F}\slashed{\partial}\psi_{A} - \eta^{\alpha\beta}j_{\alpha}A_{\beta}
\Bigr)
\label{marino}
\eeq
where $x^{0} = v_{F}t$ and $j^{\mu} = e\bar{\psi}_{A} \gamma^{\mu}\psi_{A} = e(\bar{\psi}_{A} \gamma^{0}\psi_{A},v_{F}\bar{\psi}_{A} \gamma^{i}\psi_{A})$, $i = 1, 2$. In such expressions, $\psi_{A}$ is a $2$-component Dirac field, $\bar{\psi}_{A} = \psi^{\dagger}_{A}\gamma^{0}$ is its adjoint, $F_{\mu\nu} = \partial_{\mu}\,A_{\nu} - \partial_{\nu}\,A_{\mu}$, $\gamma^{\mu}$ are rank-$2$ Dirac matrices given by $\gamma^{0} = \sigma^{3}$, $\gamma^{1} = i\sigma^{2}$, $\gamma^{2} = -i\sigma^{1}$, satisfying $\{\gamma^{\mu},\gamma^{\nu}\} = 2\eta^{\mu\nu}$, with $\sigma_{j}$ being the usual Pauli matrices. Also, $A$ denotes a flavor index, specifying the spin component and the valley to which the charge carrier belongs. Since the natural velocity in the gauge sector is that of light, whereas the one occurring in the fermionic sector is the Fermi velocity $v_{F}$, Lorentz invariance is broken. An $SU(4)$ version of this model has been recently used to study dynamical gap generation and chiral symmetry breaking in graphene~\cite{van}. 

The above action describes the interaction between a fermion field in $d_{e}$ dimensions with a gauge field in $d_{\gamma}$ dimensions, with $d_{e} < d_{\gamma}$. Specifically for our purposes $d_{\gamma} = d+1$ and $d_{e} = (d-1) + 1$. In addition, the indices run as follows: For the first term $\mu = 0, 1, \ldots, d$, and for the second term $\mu = \mu_{e} = 0,1,\ldots, (d - 1)$. For the case of graphene, $d=3$. Eq.~(\ref{marino}) can also be written as
\beq
S = \int d^{d_{\gamma}}x \left[ -\frac{1}{4}\,F_{\mu\nu} F^{\mu\nu} 
- \frac{1}{2\xi} (\partial_{\mu} A^{\mu})^2  
+  \Bigl( \bar{\psi}_{A}\,iv_{F}\slashed{\partial}\psi_{A} - \eta^{\alpha\beta}j_{\alpha}A_{\beta}
\Bigr) \delta(x^{d_{\gamma} - d_{e}}) \right]
\label{marino2}
\eeq
and then the conserved current is defined as
\beq
j^{\mu}(x) = e \, \bar{\psi}_{A} \gamma^{\mu} \psi_{A} \, 
\delta(x^{d_{\gamma} - d_{e}}), \,\,\, \mu = \mu_{e}
\label{current}
\eeq
and the other components are zero. In this work, we will be particularly interested in an alternative model; this corresponds to a vanishing space-time anisotropy and describes the IR Lorentz invariant fixed point where $v_{F} \to 1$ and the interaction is fully retarded (for a complete discussion see Ref.~\cite{Vozmediano:1994}). To consider the situation away from this fixed point, one should consider the replacement $\gamma^{i} \to v_{F} \gamma^{i}$.

By integrating out the degrees of freedom transverse to the $d_{e}$-dimensional space, one obtains the gauge propagator on the plane, which for the case of graphene reads 
\beq
D_{0\mu\nu}(p^2) = \frac{-i}{2\sqrt{p^2}}\left[\eta_{\mu\nu}-\frac{1-\xi}{2}\frac{p_\mu p_\nu}{p^2}\right].
\label{reducedpropagator}
\eeq
It is possible to introduce now a $d=3$ gauge field $\tilde{A}_\mu$ on the plane that propagates like Eq. \eqref{reducedpropagator}. The resulting theory is the RQED mentioned above, also known as Pseudo-Quantum Electrodynamics (PQED)~\cite{marino}
\beq
S = \int d^3x \left[-\frac{1}{2}\tilde{F}_{\mu\nu}\frac{1}{\sqrt{-\Box}}\tilde{F}^{\mu\nu}-\frac{1}{2\tilde{\xi}}\partial_\mu \tilde{A}^\mu\frac{1}{\sqrt{-\Box}}\partial_\nu \tilde{A}^\nu+\bar{\psi}_{A}\,iv_F\slashed{\partial}\psi_{A} - \eta^{\alpha\beta}j_{\alpha}\tilde{A}_{\beta} \right].
\label{marino3}
\eeq
Actions \eqref{marino} and \eqref{marino3} are physically equivalent. Whether one should employ one or the other depends on the situation. The mixed-dimensional \eqref{marino} is adequate for position space methods whereas action \eqref{marino3} is better suited for momentum space techniques. However for the purpose of perturbation theory in momentum space it suffices to derive the propagator \eqref{reducedpropagator} without knowledge of \eqref{marino3}.

\subsection{RQED in curved space}

In this paper we are interested in the curved-space version of the Lorentz invariant fixed-point model. That is:
\beq
S = \int d^{d_{\gamma}}x \sqrt{-g}\left[ -\frac{1}{4}\,F_{\mu\nu} F^{\mu\nu} 
- \frac{1}{2\xi} (\nabla_{\mu} A^{\mu}) \right]
+  \int d^{d_{e}}x \sqrt{-H} \bar{\psi}_{A}\,i \gamma^{\mu}(x)( \partial_{\mu} + \Omega_{\mu} + i e A_{\mu})\psi_{A}.
\label{curved-marino2}
\eeq
where $F_{\mu\nu} = \nabla_{\mu}A_{\nu} - \nabla_{\nu}A_{\mu} = \partial_{\mu}\,A_{\nu} - \partial_{\nu}\,A_{\mu}$ (the connection terms cancel), $\gamma^{\mu}(x) = e_{a}^{\ \mu}(x) \gamma^{a}$, 
$(\Omega_{\mu})^{\beta}\,_{\alpha} = (1/2)\omega_{\mu}\,^{ab}(J_{ab})^{\beta}\,_{\alpha}$, $(J_{ab})^{\beta}\,_{\alpha}$ being the Lorentz generators in spinor space, and $\omega_{\mu}\,^{a}_{\ b} = e_{b}\,^{\nu} ( -\delta^{\lambda}\,_{\nu}\partial_{\mu} + \Gamma^{\lambda}\,_{\mu\nu} ) e^{a}\,_{\lambda} $ is the spin connection, whose relation to the Christoffel connection
comes from the metricity condition: $\nabla_{\mu}e^{a}\,_{\nu} = \partial_{\mu}e^{a}\,_{\nu} - \Gamma^{\lambda}\,_{\mu\nu}e^{a}\,_{\lambda} + (\omega_{\mu})^{a}\,_{b}e^{b}\,_{\nu} = 0$. We have introduced the vielbein $e^{a}\,_{\lambda}$, which satisfies $\eta_{ab}e^{a}\,_{\mu}e^{b}\,_{\nu}=g_{\mu\nu}$. In addition, $H_{\alpha\beta}$ is the induced metric on the boundary of the space-time with metric $g_{\mu\nu}$. Besides the hypothesis of weak curvatures made in section \ref{lmsr}, the formalism presented here can be elaborated without fixing a particular form to the metric. That said, for application to the specific case of graphene, one usually considers metrics in a normal Gaussian-coordinate form, that is (in four space-time dimensions)
\beq
g_{\mu\nu} dx^{\mu} dx^{\nu} = dt^2 - dz^2 - h_{ij} dx^{i} dx^{j}
\label{graphenemetric}
\eeq
where $i,j = 1, 2$. In this case,
$$
\int d^{d_{\gamma}}x \sqrt{-g} = \int dt \int dz 
\int dx^{1} dx^{2} \, \sqrt{h} 
$$
where $h$ is the determinant of the spatial metric $h_{ij}$. Henceforth we will consider this form for the metric in the subsequent calculations.

Eq.~(\ref{curved-marino2}) can also be written in a form similar to Eq.~(\ref{marino2}), so that the conserved current will have an expression similar to~(\ref{current}). However, one can also consider an alternative form that will be useful in what follows. Define
\beq
\bar{e}^{a}\,_{\mu}(x) =
    \begin{cases}
      e^{a}\,_{\mu}(x) \delta(x^{d_{\gamma} - d_{e}}) & a,\mu = \mu_{e}\\
      0 & a,\mu = d_e, \ldots, d_{\gamma}-1.
    \end{cases} 
\eeq
In the case of graphene, $x^{d_{\gamma} - d_{e}} = z$, see Eq.~(\ref{graphenemetric}). Moreover, we consider that the extra dimensions $d_{\gamma} - d_{e}$ are all flat which justifies the usage of the standard Dirac delta function. In this way the action displays a form which closely resembles the one of the standard QED in curved space, namely
\beq
S = \int d^{d_{\gamma}}x \sqrt{-g} \left[ -\frac{1}{4}\,F_{\mu\nu} F^{\mu\nu} 
- \frac{1}{2\xi} (\nabla_{\mu} A^{\mu}) 
+  \bar{\psi}_{A}\,i \bar{\gamma}^{\mu}(x)( \partial_{\mu} + \Omega_{\mu} + i e A_{\mu})\psi_{A} \right]
\label{curved-marino3}
\eeq
where $\bar{\gamma}^{\mu}(x) = \bar{e}_{a}^{\ \mu}(x) \gamma^{a}$. Action \eqref{curved-marino3} will be the starting point of our analysis. In order to carry a one-loop analysis our first goal is to derive a curved space version of propagator \eqref{reducedpropagator}. This is done in section \ref{lmsr} where we also discuss in the end the possibility to generalize the PQED action \eqref{marino3} itself.

\subsection{Ward Identity for curved space RQED}
\label{wardcurved}

Consider the path-integral formulation of the theory, whose generating functional is given by
\beq
Z = \int D A_{\mu} D\psi D\bar{\psi} \exp\left\{i S 
+ i\int d^{d_{\gamma}}x \sqrt{-g}\left( J^{\mu} A_{\mu} + \bar{\eta}\psi + \bar{\psi} \eta \right) \right\}
\eeq
where $S$ is given by~(\ref{curved-marino3}). There should be also the contribution of the Faddeev-Popov ghost fields to the generating functional which is important in the evaluation of the one-loop effective action; since they will not play a role in our investigation, we choose to omit them for brevity.

Using functional methods, it is not difficult to exhibit the Schwinger-Dyson equation for the fermion propagator:
\beq
-i S^{-1}(x,x') = - i S^{-1}_{0}(x,x') + i \Sigma(x,x')
\eeq
where $S_{0}$ is the free curved-space counterpart of the fermion propagator and the self-energy reads
\beq
-i \Sigma(x,x') = \int d^{d_{\gamma}}z \sqrt{-g(z)} \int d^{d_{\gamma}}u \sqrt{-g(u)}
 (-i e \gamma^{\mu}(x))  i S(x,u) \bigl( -i e \Gamma^{\nu}(u,x'; z) \bigr) i G_{\mu\nu}(z,x)
\eeq
with $G_{\mu\nu}$ being the exact gauge propagator. In addition, $\Gamma^{\nu}(u,x'; z)$ is the exact three-point function with the external exact propagator removed:
\beq
\Gamma^{\nu}(u,x'; z) = \frac{\delta^{3} \Gamma}{\delta A_{\nu}(z) \delta\psi(u) \delta\bar{\psi}(x') }
\eeq
where $\Gamma$ is the proper vertex and the functional derivatives are taken with respect to the so-called classical fields. The inverse fermion propagator can also be given as
\beq
S^{-1}(x,x') = \frac{\delta^{2} \Gamma}{\delta\psi(x) \delta\bar{\psi}(x') }.
\eeq
The derivation of the Schwinger-Dyson equation for the gauge propagator follows along similar lines; one finds
\beq
-i G_{\mu\nu}^{-1}(x,x') = - i G^{-1}_{0\mu\nu}(x,x') - i \Pi_{\mu\nu}(x,x')
\eeq
where $G_{0\mu\nu}$ is the free gauge propagator in curved space. The vacuum polarization is defined as
\beq
i \Pi^{\mu\nu}(x,x') = - \int d^{d_{\gamma}}y \sqrt{-g(y)} \int d^{d_{\gamma}}y' \sqrt{-g(y')} 
\Tr \Bigl[ (-i e \gamma^{\mu}(x)) i S(x,y) \bigl( -i e \Gamma^{\nu}(y,y';x') \bigr) 
i S(y',x) \Bigr] .
\eeq
QED in curved space-time has been discussed in several places in the literature, see for instance Refs.~\cite{Shapiro06,Shapiro09,Shapiro10} and the monograph~\cite{bos}. In turn, a proof of the Ward-Takahashi identity for QED in curved space can be found, for instance, in Ref.~\cite{Panangaden:81}. In the present case we can follow a similar procedure. Namely, let $A_{\mu}$ change by 
$\nabla_{\mu}\varphi(x)$. This amounts to consider a change in $\bar{\psi}$ and $\psi$,
$$
\bar{\psi}(x) \to e^{-i e \, \varphi(x)} \bar{\psi}(x).
$$
This implies the following change in $S^{-1}(x,x')$:
\beq
\delta S^{-1}(x,x') = e \int d^{d_{\gamma}}y \sqrt{-g(y)} \, \varphi(y) \nabla_{\mu} \Gamma^{\mu}(x,x';y).
\eeq
But $\delta S^{-1}(x,x')$ can also be calculated from the transformation for in $\bar{\psi}$ and $\psi$, which gives 
\beq
\delta S^{-1}(x,x') = i e \int d^{d_{\gamma}}y \sqrt{-g(y)} \, \varphi(y) [\delta(x,y) - \delta(x',y) ] S^{-1}(x,x')
\eeq
where $\delta(x,y) = (-g(y))^{-1/2} \delta^{d_{\gamma}}(x-y)$. Comparing both expressions, one arrives at the Ward-Takahashi identity
\beq
\nabla_{\mu} \Gamma^{\mu}(x,x';y) = i [\delta(x,y) - \delta(x',y) ] S^{-1}(x,x').
\label{WT}
\eeq
Simple usage of the definition of the vacuum polarization together with Eq.~(\ref{WT}) leads us to the Ward identity in curved space:
\beq
\nabla^{x'}_{\nu} \Pi^{\mu\nu}(x,x') = 0.
\label{W}
\eeq
This derivation certainly holds for the model defined by the action~(\ref{curved-marino3}). But since this is equivalent to the action given by Eq.~(\ref{curved-marino2}), the validity of the Ward identity for RQED in curved space is hence established.

\section{Local Momentum Space Representation}\label{lmsr}

\subsection{$\text{QED}_4$}

In order to deal with the curved propagators, we employ Riemann normal coordinates (RNC) with origin at the point $x'$~\cite{Bunch:79}. This point is fixed and all other points will be in a normal neighborhood of $x'$. This means that, in the loop expressions to follow, $x$ is free to vary in a normal neighborhood of the fixed point $x'$. At the same time we make a Schwinger-DeWitt proper time expansion for the fermion and gauge propagators. Together with the RNC this leads to the so-called Local Momentum Space Representation. The usefulness of the local-momentum space representation is twofold. The first, practical reason, is that it yields to the standard momentum space techniques because only flat space-time quantities enter due to the RNC expansion. The second, most relevant and physical, is that it carries some non-perturbative information due to a partial ressumation of the scalar curvature. Here we give a qualitative overview of this approach to contextualize later discussions and comments. The detailed derivation of the fermionic and gauge propagators are defered to the appendices. See also Refs.~\cite{Barvinsky:85,Shapiro:08}.

We begin setting up the wave equations obeyed by the propagators $G^{i}_{\ j}(x,x')$~\cite{Parker:84}
\beq
\bigl[ \delta^{i}_{k} \nabla^{\mu} \nabla_{\mu} + Q^{i}_{\ k}(x) \bigr] G^{k}_{\ j}(x,x') = \vartheta \delta^{i}_{j} \delta(x,x')
\label{green}
\eeq
where the indices $i,j$ indicate any appropriate indices carried by the fields of interest (spinor or vector), 
$\vartheta = + 1$ for the gauge field and $\vartheta = -1$ for the spinor field. $Q^{i}_{\ k}(x)$ is a function with indices of the indicated type, and, as above, $\delta(x,x') = |g(x)|^{-1/2} \delta(x-x')$. Moreover, the covariant derivative in the above expression acts upon the $x$-dependence of the Green's function and is defined by
\beq
\nabla_{\mu} G^{i}_{\ j}(x,x') = \partial_{\mu} G^{i}_{\ j}(x,x') + \Gamma_{\mu}\,^{i}\,_{k}(x) G^{k}_{\ j}(x,x')
\eeq
where $\Gamma_{\mu}\,^{i}\,_{k}$ is the appropriate connection for the given spin. For the free gauge field in the Feynman gauge, Eq.~\eqref{green} is simply
\beq
\bigl[ \eta_{\mu\lambda} \Box + R_{\mu\lambda} \bigr] G^{\lambda}_{\ \nu'} = \eta_{\mu\nu'} \delta(x,x')
\label{gauge}
\eeq
so we see that $Q^{\mu}_{\ \nu} = R^{\mu}_{\ \nu}$. The free massless spinor field satisfies the equation
\beq
i \gamma^{\mu} \nabla_{\mu} S_0(x,x') = \delta(x,x').
\label{fermion1}
\eeq
However, defining $S_0(x,x') = i \gamma^{\mu} \nabla_{\mu} G(x,x')$ and using the identity~\cite{Parker:84}
$$
\gamma^{\mu} \gamma^{\nu} \nabla_{\mu} \nabla_{\nu} \Psi = \left( \Box + \frac{1}{4} R \right) \Psi
$$
where $\Psi$ is any appropriate test function, one obtains that
\beq
\left( \Box + \frac{1}{4} R \right) G(x,x') = - \delta(x,x').
\label{fermion2}
\eeq
So we observe that $Q^{i}_{\ j} = \delta^{i}_{j} R/4$, where $i,j$ are now spinor indices. Indeed, $G(x,x')$ is a bispinor.

Let us first discuss the Riemann normal coordinates expansion. In simple terms it amounts to the application of the Equivalence Principle on some point $x'$. This allows a strictly flat space-time description on $x'$ where the standard methods of field theory are valid. For points within the normal neighborhood of $x'$ we pick corrections that are polynomial in the curvature tensors and their derivatives computed at $x'$.

The Schwinger-DeWitt expansion on the other hand is done directly on the fields' propagators. It makes use of the fact that $G(x,x')$ is a transition amplitude $\langle x,s|x',0 \rangle$ evolving under a Schrödinger equation from proper time $\tau=0$ to $\tau=s$. For $x\rightarrow x'$ we fall into the domain of validity of the RNC expansion, which ultimately leads to the following fermionic and gauge propagators (see appendices)
\bea
\hspace{-20mm}
S_0(x,x')  &=& \int \frac{d^{D} k}{(2\pi)^{D}} e^{- ik y} 
\left[\frac{ \gamma^{\nu} k_{\nu} }{k^2 - M_{e}^2}
+ \frac{1}{(k^2 - M_{e}^2)^{2}} 
\left( \frac{1}{2}  R_{\nu \rho } \gamma^{\nu}  k^{\rho}
-  \frac{\gamma^{\nu} k_{\nu}}{6} R \right)
\right.
\nn\\
&+& \left.  \frac{2}{3} \frac{ \gamma^{\nu} k_{\nu} k^{\sigma} k^{\rho} R_{\rho\sigma}}{(k^2 - M_{e}^2)^3} 
+ \cdots \right]
\nn\\
\hspace{-10mm}
D_{0\mu\nu'}(x,x') &=& - 
 \int \frac{d^{d_{\gamma}} k}{(2\pi)^{d_{\gamma}}} e^{- iky} 
\left[\frac{ \eta_{\mu\nu'} }{k^2 - M_{\gamma}^2}
+ \frac{1}{(k^2 - M_{\gamma}^2)^{2}} 
\left( \frac{2}{3}  R_{\mu\nu'} - \frac{1}{6} R \eta_{\mu\nu'} \right)
\right.
\nn\\
&-& \left.  \frac{2}{3} \frac{ ( 2 R_{\mu\alpha\beta\nu'} - R_{\alpha\beta} \eta_{\mu\nu'}) 
k^{\alpha} k^{\beta}}{(k^2 - M_{\gamma}^2)^3} 
+ \cdots \right].
\label{propagatorslmsr}
\eea
In the above $M^2_e=R(x')/12$ and $M^2_\gamma=-R(x')/6$ are the result of a non-perturbative ressumation. Some comments are in order. First notice that $R(x')$ being computed at $x'$ is formally a number. Furthermore since this is a semiclassical approximation neither $M^2_e$ nor $M^2_\gamma$ are subject to renormalization. Finally we must be careful before interpreting the poles at $M^2_e$ and $M^2_\gamma$ as physical masses because for a generic curved space-time there is no unambiguous split between positive and negative frequencies to define one-particle states. For instance our general proof of the Ward Identity guarantees that there is no conflict between the parameter $M_\gamma$ and gauge invariance.

Obviously, the local-momentum space representation provides only a local approximation to the propagator. However, it should give reasonable approximate results as long as curvature effects remain weak. It is in this sense that the expression for the optical conductivity to be calculated later on is to be regarded as a high-frequency expansion.

\subsection{Reduced QED}

Up until now our discussion parallels the one for standard curved $\textrm{QED}_4$. We still need to reduce the gauge sector down to $(2+1)$ dimensions. This is a difficult task for a general curved space-time, but within the regime of validity of the local momentum space representation it can be done in the exact same fashion as in the flat space-time case. We find to first order in the Feynman gauge
\beq
D_{0\mu\nu'}(k^2) = \frac{ -i\eta_{\mu\nu'} }{2(k^2 - M_{\gamma}^2)^{1/2}}.
\eeq
This is the propagator we shall employ in the following one-loop analysis. 

In a general gauge the gauge field propagator in $\text{QED}_4$ is to first order in the local momentum space representation\footnote{Propagator \eqref{qed4lmsr} seems to break gauge invariance upon contracting with $p^\mu$ as there is a leftover proportional to $M^2_\gamma$. This is in no contradiction with our general result for gauge invariance in Section~\ref{wardcurved}. To understand the issue one must notice that the leftover is of higher order in the local momentum space expansion \eqref{propagatorslmsr}. Inspecting the corresponding higher order contribution reveals a canceling term. }
\beq
D_{0\mu\nu'}(k^2) = \frac{-i}{k^2-M^2_\gamma}\left[\eta_{\mu\nu'}-(1-\xi)\frac{k_\mu k_{\nu'}}{k^2-M^2_\gamma}\right].
\label{qed4lmsr}
\eeq
Within this approximation we find upon projection 
\beq
D_{0\mu\nu'}(k^2) = \frac{ -i }{(k^2 - M_{\gamma}^2)^{1/2}}\left[\eta_{\mu\nu'}-(1-\tilde{\xi})\frac{k_\mu k_{\nu'}}{k^2 - M_{\gamma}^2}\right].
\label{reducedpropagatorlmsr}
\eeq
It is not straightforward, if possible at all, to infer a purely $(2+1)$-dimensional action that reproduces \eqref{reducedpropagatorlmsr} in analogy to the passage from \eqref{reducedpropagator} to \eqref{marino3} - i.e. a curved space generalization to PQED. Furthermore, if indeed possible this would only be an UV limit of curved PQED. Finally we notice this is the reason for choosing the name curved RQED instead of curved PQED for the approach we adopt in this work.

\section{One-Loop Analysis}


In this work we are interested in calculating the one-loop diagrams:
\bea
i \Pi^{\mu\nu}_{1}(x,x') &=& -  \Tr \Bigl[ (-i e \gamma^{\mu}(x)) i S_{0}(x,x') (-i e \gamma^{\nu}(x')) 
i S_{0}(x',x) \Bigr]
\nn\\
-i \Sigma_{1}(x,x') &=&  (-i e \gamma^{\mu}(x))  i S_{0}(x,x') (-i e \gamma^{\nu}(x')) i D_{0\mu\nu}(x',x)
\nn\\
-i e \Gamma^{\mu}(y,y';x) &=& (-i e \gamma^{\beta}(x)) i S_{0}(x,y) (-i e \gamma^{\mu}(y)) i S_{0}(x,y')
(-i e \gamma^{\alpha}(y')) i D_{0\alpha\beta}(y',y)
\label{oneloop}
\eea
where $D_{0\mu\nu}$ are the (free) curved-space counterpart of the reduced gauge field propagators, respectively. The one-loop fermion propagator is then given by
\beq
i S(x,x') = i S_{0}(x,x') + \int d^{d_{\gamma}}z \sqrt{-g(z)} \int d^{d_{\gamma}}z' \sqrt{-g(z')}
 i S_{0}(x,z) \bigl(-i \Sigma_{1}(z,z')\bigr) iS_{0}(z',x').
\label{oneloopf}
\eeq
As standard in QFT calculations, some of such integrals are divergent, and a careful procedure of regularization and renormalization should be taken into account. As quoted above, quantum electrodynamics in curved space has been considerably discussed in the literature~\cite{Shapiro06,Shapiro09,Shapiro10,bos,Panangaden:81}. Following~\cite{Teber:2012}, we employ dimensional regularization. Loop integrals will depend on $d_{e}$ which is given as a function of suitable quantities $\epsilon_{\gamma}$ and $\epsilon_{e}$:
$$
d_{e} = 4 - 2 \epsilon_{\gamma} - 2 \epsilon_{e}.
$$
After evaluating the loop integrals for a general $d_{e}$, we employ the above expression for a fixed value of $\epsilon_{e}$, namely $\epsilon_{e}=1/2$. The associated divergences will correspond to poles in $1/\epsilon_{\gamma}$. The relation between bare and renormalized quantities follows the usual recipe,
\bea
\psi &=& Z^{1/2}_{2} \psi_{R} 
\nn\\
A &=& Z^{1/2}_{3} A_{R}
\nn\\
e &=& Z_{e} e_{R} = \frac{Z_{1}}{Z_{2} Z^{1/2}_{3}} e_{R} 
\nn\\
\Gamma^{\mu}_{R} &=& Z^{-1}_{1} \Gamma^{\mu}
\nn\\
\xi &=& Z_{3} \xi_{R}
\label{renormalized}
\eea
where the subscript $R$ means a renormalized quantity. As usual, a renormalization scale $\tilde{\mu}$ with dimensions of mass must be introduced. One then rewrites the Lagrangian density in terms of such renormalized quantities and renormalization constants that absorb all UV divergences. Use of the Ward-Takahashi identity~(\ref{WT}) leads to $Z_{1} = Z_{2}$. In the modified minimal subtraction scheme (which we adopt here) the renormalization constants take a simple form
\beq
Z_{n} = 1 + \delta Z_{n}(\alpha_{R},\epsilon_{\gamma}),\,\,\, n =1,2,3
\label{RC}
\eeq
where $\alpha_{R} = e_{R}^2/4\pi$ is the renormalized fine-structure constant and $\delta Z_{n}(\alpha_{R},\epsilon_{\gamma})$ is expanded in powers of $\alpha_{R}$ and $1/\epsilon_{\gamma}$. Taking into account such an expression for the renormalization constants, one obtains a counterterm Lagrangian density (besides a Lagrangian density written in terms of only the renormalized fields and parameters).

A theorem proved by Collins states that all counterterms are necessarily local in a flat background~\cite{Collins:1974}. An important consequence of this theorem is that a non-local contribution in the action does not get renormalized (i.e., the associated $\delta Z = 0$). This point is extensively discussed in Ref.~\cite{teber}. In the case of RQED in flat space, this implies that the beta function is zero to all orders in perturbation theory, producing thereby an explicit example of an interacting boundary conformal field theory. On the other hand, since the discussion in Ref.~\cite{Collins:1974} was based on the analysis of superficial degree of divergence of Feynman diagrams, in a general curved space, when one combines the local-momentum representation and the usual Feynman technique, one obtains that the necessary counterterms must also be covariant local expressions. This is because divergences in loops should be tantamount to local effects. However, one can argue that propagators represent correlations of fields at different space-time points, so one must be able to obtain non-local contributions. Indeed, these arise from the finite parts of the loops, but not from the UV divergences. The uncertainty principle lies underneath this split: Ultraviolet divergences must be associated with the high-energy contribution and hence emerge as short-distance (local) effects. This should also hold in curved space-times, even if the theory contains some non-local operators. That is why one should naturally expect that non-local contributions in the curved-space action will not get renormalized. In particular, this suggests that the beta function of curved-space RQED should also be zero to all orders in perturbation theory. The one-loop proof of this statement will be given in due course. 

\subsection{One-loop fermion Self-Energy}

\begin{figure}[!t]
\includegraphics[scale=0.5]{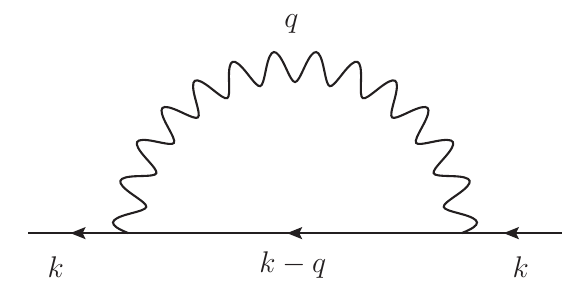}
\caption{One-loop fermion Self-Energy.}
\label{FSE}
\end{figure}

Now we turn to the task of studying the one-loop diagrams in detail. We start our discussion with the fermion self-energy. This is given by the second expression in Eq.~(\ref{oneloop}), see also Fig.~\ref{FSE}. Considering Riemann normal coordinates with origin at $x'$, one must insert into such an expression the propagators calculated in the appendices, which are given as expansions in the curvatures. Concerning the gauge propagator, and as discussed above, one should integrate out the gauge degrees of freedom transverse to the $d_{e}$-dimensional space in which the fermion lives. This amounts to consider an integration over the $d_{\gamma}-d_{e}$ bulk degrees of freedom of the gauge propagator, whose expression is derived in Appendix~\ref{B}. After specializing to $d_{e} = (2+1)$-dimensional case, one finds the following local-momentum representation for the one-loop fermionic self-energy:
\begin{equation}
   \Sigma_{1}(k,x') =  \frac{1}{2} \int \frac{d^3q}{(2\pi)^3}(-ie\gamma^\mu)
   \frac{i(\slashed{k}-\slashed{q})}{(k-q)^2-M^2_e + i\epsilon}(-ie\gamma^\nu)
   \frac{i\eta_{\mu\nu}}{(q^2-M^2_\gamma + i\epsilon)^{1/2}}.
\end{equation}
where, as defined in the appendices, $M_{e}^2 = R(x')/12$ and $M_{\gamma}^2 = - R(x')/6$. As is clear from the above expression, we are working in the Feynman gauge, $\xi =1$. Moreover, notice that we kept only the leading-order terms in the expansion in curvatures for the propagators. These are the only ones that will generate a divergence at $d_{e}=3$ and hence to a $\tilde{\mu}$ dependence. Accordingly, we also kept only the leading-order term in the expansion of the gamma matrices, so the $\gamma$'s in the above equation are just the standard flat-space gamma matrices in three dimensions. Finally, observe the introduction of the $i \epsilon$'s in the denominators of the propagators. These are necessary in order to take into account the time-ordering boundary condition.

Using that $\gamma^\mu\gamma^\alpha\gamma_\mu = -\gamma^\alpha$ and introducing usual Feynman parameters and using dimensional regularization, together with standard techniques, one finds
\begin{equation}
   \Sigma_{1}(k,x') = \frac{e^2\slashed{k}}{16\pi^{2}}\int_0^1du\sqrt{1-u}
   \left[\frac{1}{\bar{\epsilon}_{\gamma}}
   -\ln\left(\frac{\Delta - i\epsilon}{\mu^2}\right)\right]
\end{equation}
where $\Delta = \Delta(u) = uM^2_e+(1-u)M^2_\gamma-u(1-u)k^2$ and
\beq
\frac{1}{\bar{\epsilon}_{\gamma}} \equiv \frac{1}{\epsilon_{\gamma}} - \gamma_{E} + \ln 4\pi,
\eeq
$\gamma_{E}$ is Euler's constant and as asserted above $2 \epsilon_{\gamma} = 3 - d_{e}$. 

Now let us present an explicit expression for the renormalization constant $Z_{2}$. Consider Eq.~(\ref{oneloopf}). Let us employ Riemann normal coordinates with origin at $x'$. In general, the expansions for $S_{0}(x,z)$ and $\Sigma_{1}(z,z')$ will be different from the expressions given previously since it is $x'$ that is fixed and the arguments of such quantities do not contain $x'$. Then one should consider for $S_{0}(x,z)$ and $\Sigma_{1}(z,z')$ a more general momentum-space representation~\cite{Bunch:81}. Nevertheless, at leading order the results are the same. Therefore, one finds the following one-loop local-momentum representation at leading order in the expansion in curvatures
\beq
S(k,x') = S_{0}(k,x') + S_{0}(k,x') \Sigma_{1}(k,x') S_{0}(k,x').
\label{propsigma}
\eeq
Now consider the leading term in the expansion of $S_{0}(k,x')$. Since curvature effects are supposed to be sufficiently small, this can also be written as
\beq
S_{0, \textrm{leading}}(k,x') = \frac{ \gamma^{\nu} k_{\nu} }{k^2 - M_{e}^2}
= \frac{ \gamma^{\nu} k_{\nu} }{k^2} + \frac{ \gamma^{\nu} k_{\nu} }{k^4}\frac{R(x')}{12} + \cdots
\eeq
in other words, we obtain the standard local-momentum representation. Hence Eq.~(\ref{propsigma}) can be written as
\beq
i S(k,x') = \frac{i}{\slashed{k}} + \frac{i}{\slashed{k}} \bigl[ -i \Sigma_{1}(k,x') \bigr] \frac{i}{\slashed{k}} 
+ \cdots
\label{propsigm2}
\eeq
where we are focusing only on the first term in the expansion for $S_{0}$ since this is the one important in discussing the renormalization. On the other hand, since $\psi = Z^{1/2}_{2} \psi_{R}$, one obtains that 
$S = Z_{2} S_{R}$. Hence following standard renormalization procedures, one finds that
\beq
Z_{2} = 1 + \frac{2}{3} \frac{\alpha_{R}}{4\pi \bar{\epsilon}_{\gamma}} + {\cal O}(\alpha^2)
\label{Z2}
\eeq
where we have replaced $\alpha$ by $\alpha_{R}$ in such an expression (this is correct to leading order). This has the same form as in flat space~\cite{Teber:2012}. Observe also that renormalization constant $Z_{2}$ at one-loop is unaffected by space-time curvature, a result similar to the standard quantum electrodynamics in curved space-time~\cite{Panangaden:81}. Curvature terms only contribute to the finite part of the self-energy:
\bea
   \Sigma_{1\textrm{F}}(k,x') &=& - \frac{e^2\slashed{k}}{16\pi^{2}}\int_0^1du\sqrt{1-u}
    \ln\left(\frac{\Delta - i\epsilon}{\mu^2}\right)
    \nn\\
    &=& - \frac{e^2\slashed{k}}{16\pi^{2}}\int_0^1du\sqrt{u}
    \ln\left(\frac{(1-u)M^2_e + u M^2_\gamma - i\epsilon -u(1-u)k^2 }{\mu^2}\right).
    \label{finiteselfenergy}
\eea
%

\subsection{One-loop vertex correction}

\begin{figure}[!t]
\includegraphics[scale=0.5]{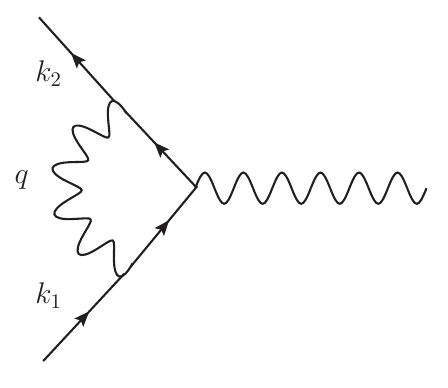}
\caption{One-loop vertex correction.}
\label{vertex}
\end{figure}

Let us now we turn our attentions to the vertex correction at one-loop order. This is given by the third expression in Eq.~(\ref{oneloop}), see Fig.~\ref{vertex}. Again considering Riemann normal coordinates with origin at $x'$, one must insert into such an expression the field propagators calculated in the appendices. By taking into account only the leading-order term of such an expansion, one finds that
\begin{equation}
 e \Gamma^{\mu}_{1}(k_{1},k_{2},x') = \frac{1}{2} \int\frac{d^3q}{(2\pi)^3} 
\frac{i\eta_{\alpha\beta}}{(q^2-M^2_\gamma + i\epsilon)^{1/2}}
(-i e \gamma^\beta)\frac{i(\slashed{k}_1+\slashed{q})}{(k_1+q)^2-M_e^2 + i\epsilon}
(-i e \gamma^\mu) \frac{i(\slashed{k}_2+\slashed{q})}{(k_2+q)^2-M_e^2 + i\epsilon}
(-i e \gamma^\alpha)
\end{equation}
where as above we have considered the reduced gauge propagator in the Feynman gauge. After introducing suitable Feynman parameters and a simple change of variables, one finds that only one of the possible terms in the numerator produces an UV divergence -- this is the one independent of $k_1$ and $k_2$. Physically we interpret it as a contribution to the charge form factor. So let us calculate the vertex function for $k_{1}=k_{2}=0$. Again following the standard procedure, one finds
\begin{equation}
\widetilde{\Gamma}^{\mu}_{1}(x') =  \frac{e^2}{32\pi^2}\int_0^1 dy dz
\frac{\theta (-y-z+1) \theta (y+z)}{\sqrt{1-y-z}}\left[ \frac{1}{\bar{\epsilon}_{\gamma}} - \frac{2}{3}
-\ln\left(\frac{\widetilde{\Delta}(y,z;0,0) - i\epsilon}{\tilde{\mu}^2}\right)\right]\gamma^\mu
\end{equation}
where
$$
\widetilde{\Delta}(y,z;k_1,k_2) 
=  (1-y-z)M^2_\gamma+(y+z)M^2_e+(yk_1+zk_2)^2-yk^2_1-zk^2_2.
$$ 
Now we must discuss the one-loop renormalization of the vertex function. This amounts to calculate the renormalization constant $Z_{1}$ at one-loop level. Proceeding as in the previous section, the vertex function up to one-loop level in the local-momentum representation can be written as
\beq
-i e  \Gamma^{\mu}(k_{1},k_{2},x') = - i e \gamma^{\mu} - i e \Gamma^{\mu}_{1}(k_{1},k_{2},x')
\eeq
where as above we considered only the leading order in the expansion in curvatures. On the other hand, the renormalized vertex function $\Gamma^{\mu}_{R}$ is given in terms of the associated bare quantity 
$\Gamma^{\mu}$ and $Z_{1}$ as
\beq
\Gamma^{\mu}_{R}(k_{1},k_{2},x') = Z^{-1}_{1} \Gamma^{\mu}(k_{1},k_{2},x')
\eeq
again in leading order in the expansion in curvatures. By using again the standard approach, one finds that
\begin{equation}
Z_{1} = 1  +  \frac{2}{3} \frac{\alpha_{R}}{4\pi \bar{\epsilon}_{\gamma}} + {\cal O}(\alpha^2).
\label{Z1}
\end{equation}
where as above we have replaced $\alpha$ by $\alpha_{R}$. A simple comparison between Eqs.~(\ref{Z1}) and~(\ref{Z2}) shows that $Z_{1} = Z_{2}$.  So we have explicitly verified the constraint between such renormalization constants at one-loop order: This result, which is a consequence of the Ward-Takahashi identity, is still valid for the curved-space version of RQED.

\subsection{One-loop vacuum polarization}

\begin{figure}[!t]
\includegraphics[scale=0.37]{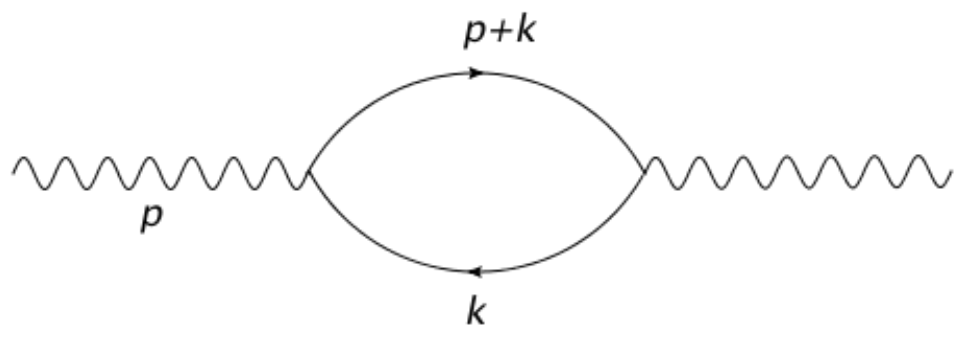}
\caption{One-loop vacuum polarization.}
\label{vacuumpol}
\end{figure}

Finally let us discuss the one-loop vacuum polarization. This is given by the first expression in Eq.~(\ref{oneloop}). See also Fig.~\ref{vacuumpol}. We will consider this calculation with more detail since we wish to explicitly check the aforementioned expectation concerning the vanishing of the one-loop beta function. Even though this is a somewhat standard calculation, we will give a step-by-step analysis of this issue, so that our conclusions are presented in a clear way. The same goes for checking the Ward identity.

Again considering Riemann normal coordinates with origin at $x'$, one must insert into such an expression the fermion propagator calculated in the Appendix~\ref{A}. By taking into account only the leading-order term of such an expansion, one finds that
\begin{equation}
  i\Pi^{\mu\nu}_1(p,x') = -e^2\int \frac{d^3k}{(2\pi)^3}
  \tr[\gamma^\mu\gamma^\alpha\gamma^\nu\gamma^\beta]\frac{(p+k)_\alpha k_\beta}{((p+k)^2-M^2_e + i\epsilon)(k^2-M^2_e+ i\epsilon)}.
\end{equation}
As above, in such an equation use is made of the flat-space version of the gamma matrices. Using properties of the traces of products of gamma matrices and introducing Feynman parameters, one finds
\begin{equation}
    i\Pi^{\mu\nu}_1(p,x') =-2e^2\int_0^1 dx \,\int \frac{d^3k}{(2\pi)^3} \frac{((1-x)p+k)^\mu(k-xp)^\nu+ \mu\leftrightarrow\nu -((1-x)p+k).(k-xp)\eta^{\mu\nu}}{(k^2-\bar{\Delta} + i\epsilon)^2},
\end{equation}
where $\bar{\Delta} = M^2_e-x(1-x)p^2$ and we have redefined $k\rightarrow k-xp$. Keeping only even terms in $k$ and considering that $k^\mu k^\nu \rightarrow k^2\eta^{\mu\nu}/3$ inside the integral, we get
\begin{equation}
   i \Pi^{\mu\nu}_1(p,x') =2e^2\int_0^1 dx\,\int\frac{d^3k}{(2\pi)^3} \frac{1}{(k^2-\bar{\Delta} + i\epsilon )^2}
   \left[\left(\frac{1}{3}k^2-x(1-x)p^2\right)\eta^{\mu\nu}+2x(1-x)p^\mu p^\nu\right].
\end{equation}
This contribution turns out to be finite. Using standard techniques to calculate the momentum integral, one obtains
\begin{equation}
    i\Pi^{\mu\nu}_1(p,x') =-\frac{ie^2}{2\pi}(p^2\eta^{\mu\nu}-p^\mu p^\nu)\int_0^1 dx\,
    \frac{x(1-x)}{\sqrt{M^2_e -i\epsilon - x(1-x)p^2}}
    +\frac{ie^2M^2_e}{4\pi}\eta^{\mu\nu}\int_0^1dx\frac{1}{\sqrt{M^2_e -i\epsilon - x(1-x)p^2}}.
\end{equation}
Apparently the Ward identity is violated by the presence of an anomalous contribution, given by the second term on the right-hand side of the above equation. However, by evaluating the $x$-integrals one finds that the transversality breaking term is actually longitudinal; more importantly, since the numerator is proportional to $M_{e}^2 = R(x')/12$, such a term is of higher order in the curvature expansion currently considered. Hence at leading order
\begin{equation}
    i\Pi^{\mu\nu}_1(p,x') = \frac{ie^2}{4\pi}(p^2\eta^{\mu\nu}-p^\mu p^\nu)
    \left[ \frac{\sqrt{M_e^2 - i\epsilon }}{p^2}
    +\frac{1}{4p}\ln\left(\frac{2\sqrt{M^2_e - i\epsilon}  - p}{2\sqrt{M^2_e - i\epsilon}  + p}\right) \right]
    \label{1loopvacpol}
\end{equation}
and the Ward identity at one-loop order is satisfied.

The most relevant upshot from this calculation is that the vacuum polarization is finite, at least at one-loop order. This means that such a contribution does not get renormalized, $\delta Z^{(1)}_{3} = 0$. This in turn implies that the beta function of the curved-space version of RQED is zero at one-loop order.

\section{Application to curved graphene layer}

As described in~\cite{Vozmediano:2010} positive or negative intrinsic curvature in graphene arises by removing or introducing sites in a given hexagonal lattice ring. These are the so-called disclination defects. Dislocation defects (pair of disclinations of opposite curvature) introduce torsion but have zero net curvature~\cite{Katanaev:92,Lorenci:2005,Lorenci:2012}. Ripples due to thermal fluctuations have also been observed~\cite{Katsnelson:2007}. In this section we describe how to apply the formalism developed so far to the case of curved graphene layers.

An idealized model of a disclination in an elastic media is obtained when the curvature is concentrated at the tip of a cone. In this case, the geometry can be described by a metric similar to the one found in the discussions of a single cosmic string~\cite{Vozmediano:2006,Lorenci:1999}. As well known, this generates a conical singularity in the curvature: The scalar curvature in this space will be proportional to a delta function~\cite{Sokolov:1977,Ribeiro:2001}. Hence it appears at first sight that disclinations cannot be encompassed in the present formalism. However, an important issue concerns the core region of the defect. Indeed, realistic models of cosmic strings, in which the space-time curvature is spread over a region of finite size, was discussed in detail in Ref.~\cite{Allen:90}. The main idea essentially consists in replacing the conical singularity by a smooth spherical cap. In such cosmological models the curvature is confined inside a cylinder, describing the interior structure of the string. Concerning a graphene sheet, the generation of a disclination by employing the usual ``cut and glue" procedure will result in a true cone. This corresponds to a point-like disclination defect and, in particular, implies the presence of a conical singularity. Notwithstanding, one must bear in mind that a membrane possesses finite elasticity, so a realistic situation must naturally go beyond the infinite-rigidity approximation. Thus, in order to have the curvature spread over some finite region one needs to take into account elastic properties. In other words, one needs to consider the graphene layer at finite elasticity~\cite{Kochetov10.1,Kochetov10.2}. For similar intriguing discussions regarding a realistic account of a physical lattice in graphene-like systems, see also Refs.~\cite{Vitagliano:18,Vitagliano:19}. In summary, a realistic picture of disclinations should take into account the finite elasticity of the graphene layer, and this amounts to considering a suitable procedure of regularization for the conical singularity (for interesting discussions regarding techniques for regularizing the conical singularity, see Refs.~\cite{Solodukhin:1995,Solodukhin:2011}). In such a situation, our formalism is expected to be fully operational and to provide sensible physical results. Other possible situations to which our formalism can be applied are those discussed in Ref.~\cite{iorio:2014}, where a quantum field theory in curved graphene was constructed. On the other hand, as well known it is possible to consider curved space-times with conical singularities but with well-behaved scalar curvatures~\cite{Oliveira-Neto:96}. In any case, around the tip of a cone (including the tip) a smooth differentiable structure is available. Indeed, a heat kernel expansion on spaces with a conical singularity can be derived~\cite{Fursaev:1994,Mooers:1999,Solodukhin:2011}, which implies that the local-momentum representation can also be used in such contexts.

There are two small modifications to be made. First of all photons, contrary to electrons, are not subjected to a curved space. Therefore we set $M_\gamma^2=0$. Notice however that the one-loop vacuum polarization \eqref{1loopvacpol} is not affected by $M^2_\gamma$ as there are no internal gauge field propagators. Importantly the Ward Identity still holds for $M_\gamma^2=0$ as only a few immaterial factors of $|g|^{1/2}$ drop out. This is confirmed by the recovery of the known UV divergences from flat graphene, see from~\cite{Vozmediano:2011}. 

In turn, we must substitute $\gamma^i$ by $v_{F}\gamma^i$, with $v_{F} \approx 1/300$. This takes into account the actual Fermi velocity of the Dirac excitations. The system \eqref{renormalized} of renormalized parameters is then complemented by
\beq
v_{F} = Z_v v_R.
\eeq 
It has been shown that the relativistic theory with $v_{F}=c=1$ is a fixed point in the infrared~\cite{Vozmediano:1994,Vozmediano:2011}. It must be kept in mind that $v$, hence also $Z_v$, enter only alongside the spatial components of the gamma matrices. This results in a slightly more involved renormalization procedure as the frequency parts of both the fermion self-energy and vertex correction are proportional to $Z_2$ and $Z_1$, whereas the momentum parts are proportional to $Z_2 Z_v$ and $Z_1 Z_v$. By virtue of the Ward Identity $Z_1=Z_2$ it is seen that the fermion wavefunction and vertex renormalize equally as usual. This suggests two independent ways to compute $Z_v$, the simplest being through the fermion self-energy.

The Feynman rules for the application of the theory to graphene for the case of retarded Coulomb interaction produce the following expressions for the fermionic and gauge-field propagator, and the photon-fermion-fermion vertex, respectively:
\bea
&iS_0(\omega_p,\mathbf{p}) = \frac{i(\gamma^0\omega_p-v_{F}\gamma^i p^i)}{\omega_p^2 - v_{F}^2\mathbf{p}^2-M_e^2v_{F}^4}\nn\\
&iD_0(\omega_p,\mathbf{p}) =\frac{1}{2} \frac{i}{\sqrt{-\omega_p^2+\mathbf{p}^2}}\nn\\
&-ie\Gamma_0^0 = -ie\gamma^0.
\eea
The free fermion propagator above has the feature that it does not modify the density of states at $x'$ because $M_e^2(x')$ is a momentum-independent constant within our framework. This readily follows from
\begin{equation}
    \rho(\omega) = -\frac{1}{\pi}\text{Im}\int d^2\mathbf{k}\, \text{Tr}\left[\frac{\gamma^0\omega-v_F\boldsymbol{\gamma}.\mathbf{k}}{\omega^2-v_F^2\mathbf{k}^2-M_e^2v_F^4}\gamma^0\right],
\end{equation}
by use of the standard identity for the principal value $P$
\begin{equation}
    P\left(\frac{1}{x\pm i \epsilon}\right) = \frac{1}{x}\mp i\pi\delta(x).
\end{equation}
Performing the integral with polar coordinates the Jacobian factor of $k$ cancels with one arising from the delta function $\delta(\omega^2-v_F^2k^2-M^2_ev_F^4)$, leading to the usual linear $\omega/v_F^2$ behavior around the Dirac points. 

\subsection{1-loop fermion self-energy}

\begin{figure}[!t]
\includegraphics[scale=0.5]{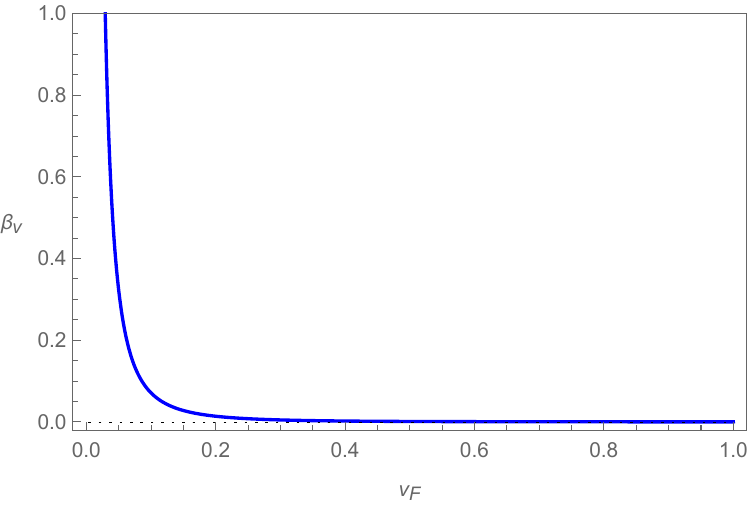}
\caption{Fermi velocity 1-loop beta function for graphene with retarded Coulomb interaction.}
\label{FermiVbeta}
\end{figure}

Let us discuss the one-loop self-energy. One finds that
\beq
-i\Sigma_1(\omega_p,\mathbf{p}) = \frac{e^2}{2}\int\frac{d^{d-1}\mathbf{k}}{(2\pi)^{d-1}}\frac{d\omega_k}{2\pi}\frac{\gamma^0(\gamma^0(\omega_k+\omega_p)-v_{F}\gamma^i(k+p)_i)\gamma^0}{((\omega_k+\omega_p)^2-v_{F}^2(\mathbf{k}+\mathbf{p})^2-M_e^2v_{F}^4)(\omega_k^2-\mathbf{k}^2)^{1/2}}.
\eeq
Having already established that the UV divergences are in general the same as in the flat model, we just state the result for the Fermi-velocity renormalization
\beq
\delta Z_v = -\frac{\alpha_g}{4\pi}\left(\frac{3}{1-v_{F}^2}-\frac{(1+2v_{F}^2)\cos^{-1}v_{F}}{v_{F}(1-v_{F}^2)^{3/2}}\right).
\eeq
which, apart from a constant factor proportional to the square of the Fermi velocity (coming from the current density interaction of the vertex we have dropped), recovers the results from Ref.~\cite{Vozmediano:1994}, to which we refer the reader for a detailed computation. The Fermi velocity beta function $\beta_v$ is shown in Fig.~\ref{FermiVbeta}. The crucial point here is that, according to our model, the relativistic fixed point achieved for $v_{F} \rightarrow 1$ is predicted to survive in the presence of disclination-induced curvature. Here $\alpha_g = e^2/(4 \pi v_F)$.
As for the finite part of the self energy, we are particularly interested in the imaginary part of $\Sigma_1^F$ as it translates to the scattering time among the charge carriers in graphene due to the electromagnetic interaction in the presence of curvature. The local $\mathbf{p}\rightarrow 0$ limit is relevant when considering level-broadening effects on the conductivity. Hence 
\beq
\Sigma_{1}^{F}(\omega_p) = -\frac{\alpha_g}{4\pi}\gamma^0\omega_p\int_0^1 dx \frac{\sqrt{1-x}}{1-x(1-v_{F}^2)}\log\left(\frac{\bar{\mu}^2}{x\left(M_e^2v_{F}^4-(1-x)\omega_p^2\right)-i\epsilon}\right)
\eeq
There are now two possible cases to consider, namely positive or negative Ricci scalar. For positive Ricci scalar, i.e. $M^2_e>0$, we obtain, for the scattering time:
\beq
\tau^{-1}_{+}(z) = \frac{\alpha_g}{4}M_e v_{F}^2 z \int_0^1 dx \frac{\sqrt{1-x}}{1-x(1-v_{F}^2)}\theta((1-x)z^2-1),   
\label{ST}
\eeq
where $z^2=\omega_p^2/M^2_e v_{F}^4$. This integrates to
\beq\label{scatteringtime}
\tau^{-1}_{+}(z) = \begin{cases} 
0, & z\leq1 \\
\frac{\alpha_g}{4}M_e v_{F}^2 z
\left(\frac{2}{1-v_{F}^2}\left(1-\frac{1}{z}\right)+\frac{2}{(1-v_{F}^2)^{3/2}}\left(\cot^{-1}\left(\frac{v_{F} z}{\sqrt{1-v_{F}^2}}\right)-\cos^{-1}v_{F}\right)\right), & z>1.
\end{cases}
\eeq
For negative Ricci scalar, i.e. $M^2_e<0$, the self-energy always acquires an imaginary part. In this case, the scattering time is given by
\beq
\tau^{-1}_{-}(z) = \frac{\alpha_g}{4}M_e v_{F}^2 z\left(\frac{2}{1-v_{F}^2}
-\frac{2 v_{F}\cos^{-1}v_{F}}{(1-v_{F}^2)^{3/2}}\right).
\eeq
From the imaginary part of the self-energy, one can use standard dispersion relations to calculate the real part and, as a result, one is able to evaluate explicitly the quasiparticle residue at the Fermi energy. As will be argued in due course, the quantity $M_e v_{F}^2$ for positive Ricci scalar may play the role of an effective chemical potential. Using this as the value of the Fermi energy in the present context, one can easily show that the quasiparticle residue asymptotically should acquire a non-zero value at the Fermi energy in the case $M_e^2 > 0$. All such results concerning the self-energy should be compared with the ones of Ref.~\cite{Vozmediano:2011,Sarma:07}.

\subsection{1-loop vertex correction}

Now let us consider the one-loop vertex correction at zero external momenta. This is given by
\beq
-ie\Gamma^\mu_1(0,0) = \frac{e^3}{2}\int\frac{d^{d-1}\mathbf{k}}{(2\pi)^{d-1}}\frac{d\omega_k}{2\pi}\frac{\gamma^0\gamma^\alpha\gamma^\mu\gamma^\beta\gamma^0k_\alpha k_\beta}{(\omega_k^2-v_{F}^2\mathbf{k}^2)^2(-\omega_k^2+\mathbf{k}^2)^{1/2}}.
\eeq
It is straightforward to check that the UV divergences match those of $-i\Sigma_1$. We once more refer to Ref.~\cite{Vozmediano:1994} for the details. The finite parts of the time and spatial components read
\beq
\Gamma^{0,F}_1\gamma^0 = -\frac{\alpha\gamma^0}{8\pi}\int_0^1 dx \frac{x}{\sqrt{1-x}}\left(\frac{1}{1-x(1-v_{F}^2)}-\frac{2v_{F}^2}{(1-x(1-v_{F}^2))^2}\right)\log\left(\frac{(1-x(1-v_{F}^2))\bar{\mu}^2}{xM_e^2v_{F}^4}\right),
\label{G0F}
\eeq
and
\beq
v_{F}\Gamma^{i,F}_1\gamma^i = -\frac{\alpha v_{F}\gamma^i}{8\pi}\int_0^1 dx\frac{x}{\sqrt{1-x}}\left(\frac{1}{1-x(1-v_{F}^2)}\log\left(\frac{(1-x(1-v_{F}^2))\bar{\mu}^2}{xM_e^2v_{F}^4}\right)+\frac{v_{F}^2}{(1-x(1-v_{F}^2)^2}\right).
\label{GiF}
\eeq
These allow us to define a suitable effective Fermi velocity:
\beq
\frac{1}{v_{\text{eff}}}=\frac{1}{v_F}\left(1+\frac{(\Gamma^{0,F}_1)^3}{\Gamma^{i,F}_1}\right).
\eeq
At the point $\bar{\mu}^2=M_e^2 v_F^4$ the correction leads to a higher effective Fermi velocity in accordance with expectation from the running in Fig.~\ref{FermiVbeta}
\begin{equation}\label{vFeff}
    v_{\text{eff}} \approx 1.0072 v_F.
\end{equation}
In Ref.~\cite{Vozmediano:2007} it was shown that the effect of curvature on the electronic properties of a graphene sheet leads to a decrease in the Fermi velocity in comparison with the free velocity. On the other hand, electron-electron interactions tend to increase the Fermi velocity. In our model, we see that both effects seem to be important, and their combination contribute decisively to a slight increase in $v_F$. We remark that this conclusion was obtained for the choice $\bar{\mu}^2=M_e^2v_F^4$, so one must be very careful with possibly naive physical interpretations. In particular, this shows once more the importance of considering a finite elasticity for the graphene layer -- for ideal disclinations, $M_e^2$ would present a sharp singular behavior, as discussed above, and this choice as a renormalization point would become problematic. A renormalization-group treatment would be most welcome here. This would indeed be interesting to explore, and we hope to consider this calculation in the future.

\subsection{Higher-frequency behavior of the optical conductivity}

\begin{figure}
\centering
\begin{subfigure}[b]{0.45\textwidth}
\centering
\includegraphics[scale=0.58]{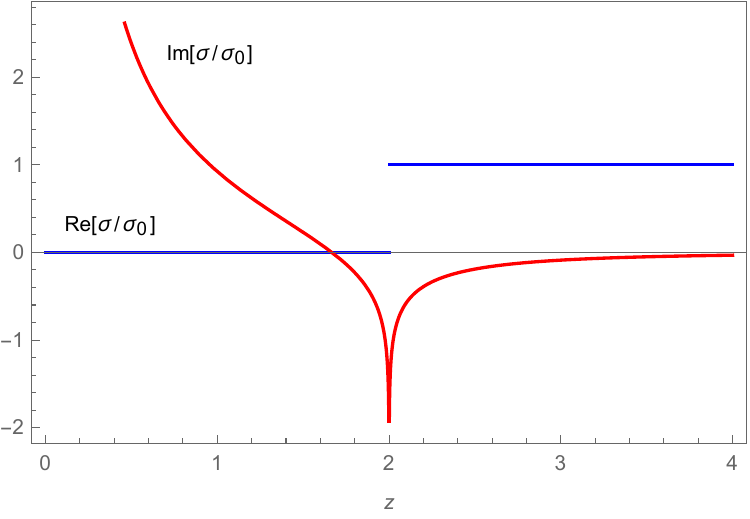}
\caption{Real and imaginary parts of $\sigma^{jj}(z,x')$ normalized to $\sigma_0$ without broadening effects.}
\label{Conductivitynobroad}
\end{subfigure}
\hfill
\begin{subfigure}[b]{0.45\textwidth}
\centering
\includegraphics[scale=0.25]{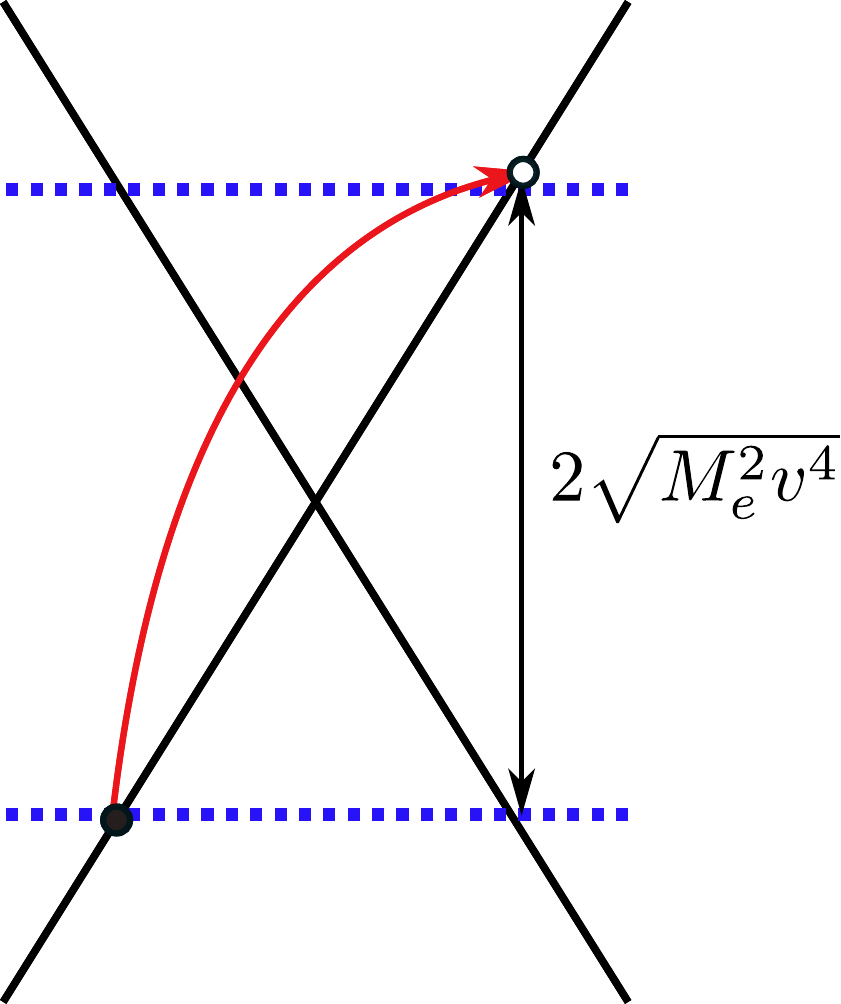}
    \caption{Origin of conductivity jump under a finite chemical potential (blue line).}
    \label{chemicalpotential}
\end{subfigure}
\caption{Non-interacting conductivity in graphene with a finite chemical potential.}
\end{figure}

As an application of the above results, let us determine the high-frequency behavior of the optical conductivity in the presence of curvature effects in graphene by using the Kubo formula, which describes the linear response to a static external electric field. In real time, it is given by
\beq
\sigma^{ik} = \lim_{{\bf p} \to 0} i \frac{\langle j^{i} j^{k} \rangle}{\omega + i\epsilon }
\eeq
where the current correlation function is meant to contain only one-particle irreducible (1PI) diagrams. A simple analysis shows that~\cite{marino2}
\beq
\langle j_{\mu} j_{\nu} \rangle_{\textrm{1PI}} = \Pi_{\mu\nu}
\eeq
where $\Pi_{\mu\nu}$ is the vacuum polarization tensor of the electromagnetic field. The optical conductivity is then given by
\begin{equation}
    \sigma^{jk}(\omega) = \lim_{\mathbf{p}\rightarrow 0} \frac{i\Pi^{jk}}{\omega + i\epsilon }.
\end{equation}
To derive the optical conductivity from the above formula, one must change the boundary conditions employed so far. This amounts to considering the various Green functions appearing in Eq.~(\ref{oneloop}) with retarded boundary conditions. In this case the loop integrals in the vacuum polarization are to be calculated using the in-in formalism, see for instance Ref.~\cite{Donoghue:2014yha}. The result has the same functional dependence, but with a different $i\epsilon $ prescription: 
$$
q^{0} \to q^0+i\epsilon.
$$
The one-loop vacuum polarization is then given by
\begin{equation}
    i\Pi^{\mu\nu}_1(p,x') = \frac{ie^2}{4\pi}(p^2\eta^{\mu\nu}-p^\mu p^\nu)
    \left[ \frac{\sqrt{M_e^2v_{F}^4}}{p^2}
    +\frac{1}{4p}\ln\left(\frac{2\sqrt{M^2_ev_{F}^4}  - p}{2\sqrt{M^2_ev_{F}^4}  + p}\right) \right], 
    \,\,\, p^{\mu} = (p^{0} +i\epsilon, {\bf p}).
\end{equation}
Geometrically it is perfectly plausible for $M_e^2$ to be either negative or positive. Both possibilities seem to lead to qualitatively different behavior due to extra factors of $i$ arising for $M_e^2<0$. In the following we will focus mostly on the positive scalar-curvature case where the physics is clearer, and we give only a brief discussion on the negative case at the end of this section. With that in mind, we combine our results to obtain the high-frequency behavior of the optical conductivity:
\begin{equation}\label{conductivity}
\sigma^{jk}(z,x') = \frac{e^2}{4}\left[\frac{4}{\pi}\frac{i}{z + i\epsilon } +1+\frac{i}{\pi}\ln\left(\frac{z+ i\epsilon -2}{z+ i\epsilon +2}\right)\right]\eta^{jk}.
\end{equation}
Observe that $\sigma^{jk}$ a function of the ratio $z=\omega/\sqrt{M^2_e v_{F}^4}$. The real and imaginary parts of $\sigma^{jk}$ are given by, for $z \neq 0:$
\begin{eqnarray}
\textrm{Re}[\sigma^{jk}(z,x')] &=& \frac{e^2}{4} \theta(z-2) \eta^{jk}
\nonumber\\
\textrm{Im}[\sigma^{jk}(z,x')] &=& \frac{e^2}{4\pi}
\left( \frac{4}{z} - \ln\left|\frac{z + 2}{z-2}\right| \right) \eta^{jk} .
\end{eqnarray}
The conductivity for the case $M_e^2 > 0$ is depicted in Fig.~\ref{Conductivitynobroad}. In this way, we recover the results presented in Refs.~\cite{KM:2016,KMP:2017} for zero temperature and zero mass gap, but finite chemical potential in the local limit. Remarkably, such equations also show that 
$\sqrt{M^2_e v_{F}^4}$ cannot play the role of a mass gap in the expression for the optical conductivity. Rather, our results, combined with those obtained in Refs.~\cite{KM:2016,KMP:2017}, seem to suggest an effective chemical-potential interpretation for $\sqrt{M^2_e v_{F}^4}$, at least as far as the optical conductivity is concerned. At the moment we do not have more elements to argue in favor of the general validity of this interpretation. In any case, this allows us to understand the first term in Eq.~(\ref{conductivity}) as due to intraband transitions, and the remaining as the interband contribution. The latter is just the minimal graphene conductivity $\sigma_0 = e^2/4$ for $z>2$. The absence of interband transitions for $z<2$ is due to the kinematics of momentum conservation of chiral fermions as illustrated in Fig.~\ref{chemicalpotential}. Even though the validity of the local momentum representation translates to high-frequency regime, our result seems to work for all $z$ given the identification $\sqrt{M_e^2 v_{F}^4}=\mu$. 

\begin{figure}[t]
    \centering
    \includegraphics[scale=0.58]{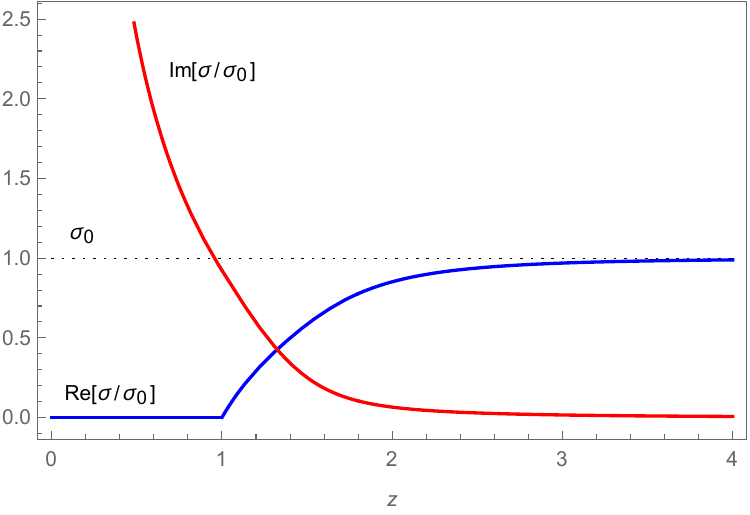}
    \caption{Real and imaginary parts of optical conductivity normalized to $\sigma_0$ with broadening effects for positive Ricci curvature scalar. Dotted line shows that the minimum conductivity $\sigma_0$ is approached asymptotically.}
    \label{ConductivityWithBroad}
\end{figure}

If one wishes to include curvature effects of level broadening due to scattering of the fermion, then one should replace $i \epsilon$ by $\tau_{+}^{-1}(z)$ in the expression of the optical conductivity. One obtains
\begin{equation}\label{conductivity2}
\sigma^{jk}(z,x') = \frac{e^2}{4}\left[
\frac{4}{\pi}\frac{i}{z + i\tau_{+}^{-1}(z)} 
+1+\frac{i}{\pi}\ln\left(\frac{z + i \tau_{+}^{-1}(z) -2}{z+ i\tau_{+}^{-1}(z) +2}\right)\right]\eta^{jk}.
\end{equation}
If $\text{Im}[\Sigma_{1}^F(\omega)]$ is small, we can approximate it as a constant value, which results in a constant $\tau_{+}^{-1}$. This implies that in this case this expression can also be obtained by employing resummed fermionic propagators in the calculation of the vacuum polarization. The result will resemble a simple one-loop calculation, even though higher-order corrections are being taken into account with the usage of dressed propagators. This is somewhat reminiscent of the standard discussion on unstable particles in high-energy scattering amplitudes within the narrow-width approximation. In the context of condensed-matter settings, a vanishingly small imaginary part of the self-energy (around the Fermi surface) implies that the criterion for the Fermi-Landau liquid theory is fully justified.

Let us first consider the full frequency dependence of $\tau_{+}^{-1}$. When $M_e^2>0$ we see from Fig.~\ref{ConductivityWithBroad} that there is no longer a jump on the real part of the conductivity at $z=2$. Instead, the conductivity starts to increase smoothly at $z=1$. Accordingly the imaginary part of $\sigma(z)$ is also smoothened at $z=2$, as dictated by the Kramers-Kronig relations. For $z\rightarrow\infty$ we still recover $\sigma_0$. Eq.~\eqref{conductivity2} is similar to the one found in Ref.~\cite{Ando:2002}, except for the fact that here the scattering time given by Eq.~\eqref{scatteringtime} kicks in only at $z=1$. Indeed, these authors considered $\tau_{+}^{-1}(z)$ at $z=1$ in the first term and 
$\tau_{+}^{-1}(z) \to \tau_{+}^{-1}(z/2)$ in the logarithmic term (in our notation) whereas we have considered the full frequency dependence of $\tau_{+}^{-1}$. The reason for this difference in the approaches lies in the fact that the energy dependence of the scattering time in both models behave differently at the Fermi energy, as just mentioned, since they describe different physical situations. Nevertheless, as shown by these authors, the scaling of the dynamical conductivity leads to a singular jump at $\omega = 0$ for zero Fermi energy, a result that clearly resembles the behavior of $\textrm{Im}[\sigma]$ in our model.

One may consider the conductivity for a fixed value of $\tau_{+}^{-1}$, somewhat partially similar to what was undertaken in Ref.~\cite{Ando:2002}. We explore this situation for the case in which $\text{Im}[\Sigma_{1}^F(\omega)]$ is small so that $\tau_{+}^{-1}$ can be taken to be approximately constant. This will take place near the Fermi energy.  As an illustration, let us quote our result for a matching scale of $z=z_0$, $z_0 \gtrsim 1$, for the scattering time (Fermi energy amounts to choosing $z_0 = 1$):
\begin{equation}\label{conductivity3}
\sigma^{jk}(z,x') = \frac{e^2}{4}\left[
\frac{4}{\pi}\frac{i}{z + i\tau_{+}^{-1}(z_0)} 
+1+\frac{i}{\pi}\ln\left(\frac{z + i \tau_{+}^{-1}(z_0) -2}{z+ i\tau_{+}^{-1}(z_0) +2}\right)\right]\eta^{jk}.
\end{equation}
It is easy to see that there is an enhancement in the conductivity for $z \geq 2$:
\beq
\sigma_{0} \to \sigma_{0} 
+ \frac{e^2}{\pi} \frac{\tau_{+}^{-1}(z_0)}{z^2 + \tau_{+}^{-2}(z_0)} .
\eeq
For $z<2$ the intraband contribution produces a positive contribution to the real part of the optical conductivity, whereas the log yields a (constant) negative contribution. However, for $z \to 0$, the intraband transition is the dominant term, and a positive contribution remains. In order to confirm this analysis we would have to calculate the optical conductivity for all regimes of frequency which would mean going beyond the large-momentum expansion used above for the propagators. We do not have a clear evaluation of this physics, but at least the conclusion seems indeed to be that curvature effects should contribute positively to the conductivity of graphene. This is in accordance with the arguments and expectations of Ref.~\cite{marino2}.

\begin{figure}[b]
\centering
\begin{subfigure}[b]{0.45\textwidth}
\centering
\includegraphics[scale=0.58]{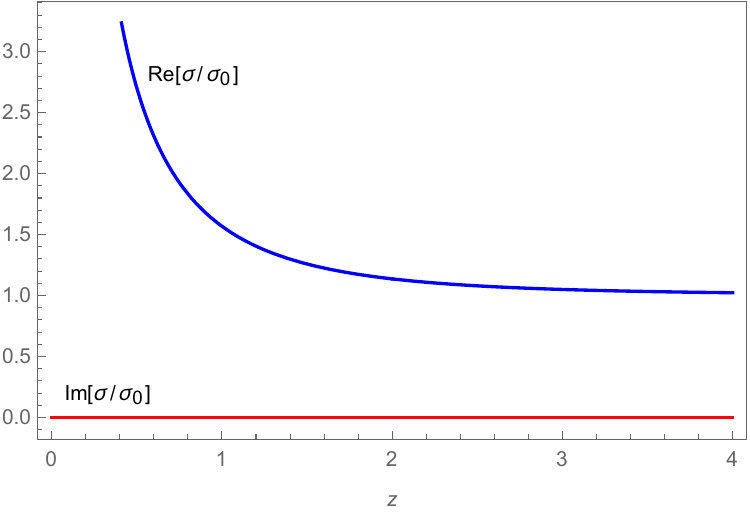}
\caption{Real and imaginary parts of non-interacting optical conductivity normalized to $\sigma_0$.}
\label{ConductivitynobroadN}
\end{subfigure}
\hfill
\begin{subfigure}[b]{0.45\textwidth}
\centering
\includegraphics[scale=0.58]{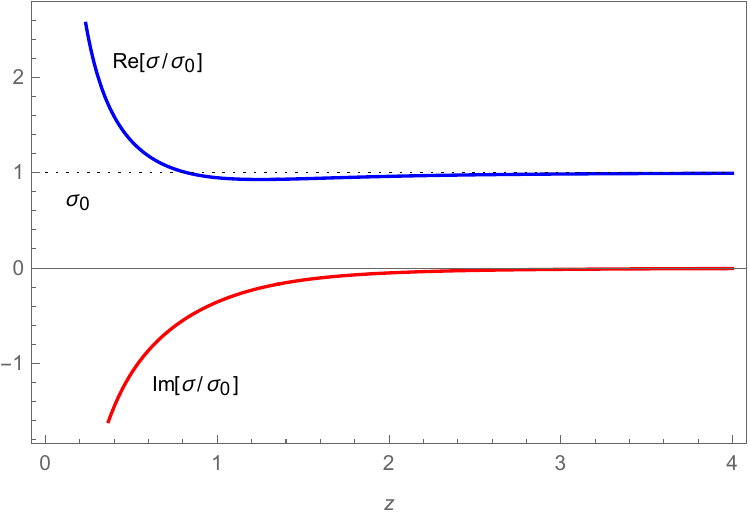}
\caption{Real and imaginary parts of optical conductivity normalized to $\sigma_0$ with broadening effects.}
\label{ConductivityWithBroadN}
\end{subfigure}
\caption{Conductivity in graphene for negative Ricci scalar.}
\end{figure}

Let us now turn our attentions to the $M_e^2 < 0$ case. The optical conductivity reads now
\beq
\sigma^{jk}(z,x') = \frac{e^2}{2}\left[\frac{4}{\pi}\frac{1}{z+i\tau_{-}^{-1}(z)}+1+\frac{i}{\pi}\ln\left(\frac{z+i\tau_{-}^{-1}+2i}{z+i\tau_{-}^{-1}-2i}\right)\right] .
\label{conductivityN}
\eeq
Fig.~\ref{ConductivitynobroadN} describes the non-interacting optical conductivity ($\tau_{-}^{-1}=0$ above). Here the model seems to run into trouble with the Kramers-Kronig relations as pointed out by the vanishing of the imaginary component. In comparison with Eq.~\eqref{conductivity}, we note the source of its imaginary component is solely due to the first term, i.e., the intraband transitions. For $M_e^2<0$ (and $\tau_{-}^{-1} = 0$) this term becomes purely real. Inclusion of broadening effects seems to lift the problem as shown in Fig.~\ref{ConductivityWithBroadN}. Here the real component also assumes a form similar to Ref.~\cite{Ando:2002} although it always stays very close to $\sigma_0$ after it crosses it from the above.

\section{Conclusions}

The primary aim of this work was to develop a formalism to study the curved-space RQED by employing the local momentum representation. Then we applied the model, with slight modifications, to graphene. In particular the optical conductivity was computed to one-loop and at leading adiabatic order surprisingly revealing the appearance of an effective chemical potential for the positive Ricci scalar case. Importantly, this effect is non-perturbative as it stems from the partial ressumation of the Ricci scalar. Furthermore, we demonstrated how the combined effect of intrinsic curvature of the graphene sheet and electron-electron interactions as described by the curved-space RQED could affect the optical conductivity. In summary, by comparing the outcomes of the present paper with the ones in the existing literature, we have showed that the curved-space RQED as a model for describing transport properties of curved graphene layers is able to (re)produce sensible physical results.

There are many open questions outside our scope that are nonetheless of great importance. Most obvious is developing curved space RQED beyond the approximations presented here. Within our approach it would also be interesting to study the trace anomaly and conformal invariance of the model. Research into possible holographic models (both for flat and curved RQED) would be most welcome for providing a tool into the non-perturbative regime. A two-loop analysis is also desirable specially for a more rigorous account of electron-electron interaction contributions to the optical conductivity. Additionally a computation of the global conductivity $\sigma(\omega)$ from the local $\sigma(\omega,x')$ by a disorder averaging treatment of $M_e^2(x')$ is expected to accurately model real samples. On the other hand, a non-trivial interesting generalization of our work could include torsion~\cite{Iorio18,Ciappina20}. We hope to access these issues in future works.

\section*{Acknowledgements} 

We thank T. Micklitz for invaluable discussions, in particular for bringing to our attention the possibility of applying techniques from quantum field theory in curved space in approaching the problem of the optical conductivity in graphene. We also thank A. Iorio and P. Pais for useful comments and discussions. The work of GM has been partially supported by Conselho Nacional de Desenvolvimento Cient\'ifico e Tecnol\'ogico -- CNPq under grant 310291/2018-6, and Funda\c{c}\~ao Carlos Chagas Filho de Amparo \`a Pesquisa do Estado do Rio de Janeiro -- FAPERJ under grant E-26/202.725/2018.

\appendix

\section{Riemann normal coordinates expansion}

The construction of Riemann normal coordinates about some point $x'$ in the manifold goes as follows. On $x'$ it is possible to make $g_{\mu\nu}(x')=\eta_{\mu\nu}(x')$ along with $\Gamma^\alpha_{\ \mu\nu}(x')=0$. Now suppose that points $x$ in the neighborhood of $x'$ can be reached by a unique geodesic starting from $x'$. This is the so-called normal neighborhood of $x'$. We can make use of the tangent vectors to the geodesics to introduce a normal coordinate system $X^\mu$ with origin at $x'$ such that
\beq
\frac{d^2X^\alpha}{d\lambda^2} = 0
\eeq
along any geodesic, with $\lambda$ some affine parameter describing the geodesic. By expanding with respect to these coordinates one finds that~\cite{Parker:84,Toms:14}
\bea
g_{\mu\nu}(x) &=& \eta_{\mu\nu} - \frac{1}{3} R_{\mu\rho\sigma\nu}(x') X^{\rho} X^{\sigma} + \cdots
\nn\\
(-g(x))^{1/2} &=& 1 + \frac{1}{6} R_{\mu\nu}(x') X^{\mu} X^{\nu} + \cdots
\nn\\
\Gamma_{\mu}\,^{i}\,_{j}(x) &=& - \frac{1}{4} R_{\mu\rho ab}(x') (J^{ab})^{i}_{\ j} X^{\rho} + \cdots
\nn\\
Q^{i}_{\ j}(x) &=& Q^{i}_{\ j}(x') + \cdots
\nn\\
e_{a}^{\ \mu}(x) &=& e_{a}^{\ \nu}(x') \left( \delta^{\mu}_{\nu} 
+ \frac{1}{6} R_{\nu\alpha}\,^{\mu}\,_{\beta}(x') X^{\alpha} X^{\beta} \right) + \cdots
\label{expansion1}
\eea
where only the lowest-order terms are retained. Here $R_{\mu\rho ab}$ is the Riemann curvature tensor with two vielbein indices and $J^{ab}$ is the Lorentz generator for the representation appropriate to the field under consideration. Also $Q^i_{\ j}$ is a quantity proportional to the curvature. Let us derive the expansion for the spin connection. From Eq.~(\ref{expansion1}), one finds
\beq
\omega_{\mu a b} =  - \frac{1}{2}  R_{\mu \rho a b }(x')  X^{\rho} 
\eeq
where we used the cyclicity property of the Riemann tensor. Hence
\beq
\Omega_{\mu} = \frac{1}{2} \omega_{\mu a b} J^{ab} = - \frac{1}{4}  R_{\mu \rho a b }(x')  X^{\rho} J^{ab}
= - \frac{1}{8}  R_{\mu \rho a b }(x')  \gamma^{a} \gamma^{b} X^{\rho} 
\eeq
which implies that
\beq
\gamma^{\mu}(x) \nabla_{\mu} = \gamma^{a} e_{a}^{\ \mu}(x) (\partial_{\mu} + \Omega_{\mu})
= \gamma^{\nu} (x') \left(  \partial_{\nu}
+ \frac{1}{6} R^{\mu}_{\ \alpha\nu\beta}(x') X^{\alpha} X^{\beta} \partial_{\mu} 
- \frac{1}{8}  R_{a b \nu \rho }(x')  \gamma^{a} \gamma^{b} X^{\rho}
\right).
\eeq
However, using the anticommutation relations for the gamma matrices and again the cyclicity property of the Riemann tensor, one finds that
$$
R_{a b c \rho } \gamma^{c} \gamma^{a} \gamma^{b} = 2 R_{a \rho }  \gamma^{a}
$$
which yields
\beq
\gamma^{\mu}(x) \nabla_{\mu} 
= \gamma^{\nu} (x') \left(  \partial_{\nu}
+ \frac{1}{6} R^{\mu}_{\ \alpha\nu\beta}(x') X^{\alpha} X^{\beta} \partial_{\mu} 
- \frac{1}{4}  R_{\nu \rho }(x')  X^{\rho} \right).
\eeq
%

\section{Local-momentum representation of the fermionic propagator}
\label{A}

In this Appendix we consider the local-momentum representation for the fermion propagator. The standard representation has been extensively discussed in the literature, see for instance Refs.~\cite{Parker:84,Inagaki:1993ya,Toms18}. Tipically, since curvature effects are small, we will be interested only in the leading terms in the Riemann curvature. But for the moment we will keep our discussion as general as possible. In principle, we could follow the same steps outlined above. There is, however, another alternative form of proper-time expansion for propagators in curved space-time which could be useful here. It is based on a partial resummation of the above series~\cite{Jack:85}. Consider Eq.~(\ref{green}) with $\vartheta = -1$. One can write the Green's function as
\beq
G(x,x') = -i \int_{0}^{\infty} ds \langle x,s | x',0 \rangle 
\eeq
where we omitted matrix indices, and the kernel $\langle x,s | x',0 \rangle$ has a Schwinger-DeWitt expansion given by~\cite{DeWitt}
\bea
\langle x,s | x',0 \rangle &=& i (4\pi i s)^{-d/2} e^{i \sigma(x,x')/2s} \Delta_{\textrm{VM}}^{1/2}(x,x')
F(x,x'; is)
\nn\\
F(x,x'; is) &=& {\bf 1} + \sum_{j=1}^{\infty} (is)^{j} f_{j}(x,x')
\eea
where $2 \sigma(x,x')$ is the square of the proper arc length along the geodesic from $x'$ to $x$ and 
$\Delta_{\textrm{VM}}(x,x')$ is the Van Vleck-Morette determinant defined by~\cite{VM}
\beq
\Delta_{\textrm{VM}}(x,x') = - |g(x)|^{-1/2} |g(x')|^{-1/2} 
\det\left[ -\frac{\partial^2 \sigma(x,x')}{\partial x^{\mu} \partial^{'\nu}} \right].
\eeq
In turn, such an expansion can be rewritten in the form
\bea
\langle x,s | x',0 \rangle &=& i (4\pi i s)^{-d/2} e^{i \sigma(x,x')/2s} \Delta_{\textrm{VM}}^{1/2}(x,x')
\bar{F}(x,x'; is) e^{-is \left[ Q(x') - \frac{1}{6} R(x') \right]}
\nn\\
\bar{F}(x,x'; is) &=& {\bf 1} + \sum_{j=1}^{\infty} (is)^{j} \bar{f}_{j}(x,x')
\eea
an assertion which was proved in Ref.~\cite{Jack:85}. The coefficients $\bar{f}_{j}(x,x')$ are $R$ independent to all orders, but generically depend on the Ricci curvature and the Riemann tensor and their powers and derivatives. In addition, we stress that such coefficients in the fermionic case should be envisaged as bispinors; hence to perform properly the above expansion one should form the contraction between such bispinors with the bispinor of parallel displacement $\boldsymbol{\sigma}(x,x')$. It can be proved that $\boldsymbol{\sigma}(x,x') = \bar{f}_{0}(x,x') = {\bf 1}$~\cite{Bunch:79}.

The term $e^{-is \left[ Q(x') - \frac{1}{6} R(x') {\bf 1} \right]}$ should be defined as a  a formal matrix power series (${\bf 1}$ is the unit spinor, in the case of fermions). A straightforward calculation yields
\bea
\bar{F}(x,x'; is) e^{-is \left[ Q(x') - \frac{1}{6} R(x') {\bf 1} \right]} &=&
{\bf 1} + (is) \left( \bar{f}_{1}(x,x') - A(x') \right) 
\nn\\
&+& (is)^{2} \left( \bar{f}_{2}(x,x') + \frac{1}{2} A^{2}(x') - \bar{f}_{1}(x,x') A(x') \right)
+ \cdots
\eea
where $A(x') = Q(x') - R(x')/6 $. Since such expansions should be equal, one finds that
\bea
\bar{f}_{1}(x,x') &=& f_{1}(x,x') + A(x')  
\nn\\
\bar{f}_{2}(x,x') &=& f_{2}(x,x') - \frac{1}{2} A^{2}(x') + ( f_{1}(x,x') + A(x') ) A(x')
\eea
and so on. On the other hand, with Riemann normal coordinates $y^{\mu}$ for the point $x$ with origin at the point $x'$, one has that
$$
f_{1}(x,x') = f_{1}(x') + f_{1\alpha}(x') y^{\alpha} + f_{1\alpha\beta}(x') y^{\alpha} y^{\beta} + {\cal O}(y^3)
$$
where an expansion about the point $x'$ was considered. The coefficients $f_{j\alpha\beta\cdots}$ are all proportional to derivatives of the $f_{j}$ evaluated at the origin of the Riemann normal coordinates (i.e., at $x'$). The coefficients $f_{j}$ have been calculated in the literature~\cite{Bunch:79}. In particular,  
$\bar{f}_{1} = 0$.

Now use the fact that, in Riemann normal coordinates about $x'$, $\Delta_{\textrm{VM}}(x,x') = |g(x)|^{-1/2}$, together with the results 
\beq
\int \frac{d^{D} k}{(2\pi)^{D}} e^{-is(-k^2 + m^2) - iky} = 
i (4\pi i s)^{-d/2} e^{i \sigma(x,x')/2s} e^{-is m^{2}}
\label{int1}
\eeq
where $\sigma(x,x') = - y_{\alpha} y^{\alpha}/2$, and
$$
\int_{0}^{\infty} ids \, e^{-is(-k^2 + m^2) } = \frac{1}{-k^2 + m^2}
$$
to obtain that
\beq
G(x,x') = \Delta_{\textrm{VM}}^{1/2}(x,x') \int \frac{d^{D} k}{(2\pi)^{D}} e^{- iky} 
\bar{F}\left( x,x'; -\frac{\partial}{\partial m^2} \right)
\frac{1}{k^2 - m^2}
\eeq
where $m^2 = Q(x') - R(x')/6 $ ($Q$ now is just a function). Here $D = d_{e}$ for the case of the fermionic propagator. We also consider the replacement
\beq
y^{\alpha} \to i \frac{\partial}{\partial k_{\alpha}}
\label{int2}
\eeq
in the above expression. 

Now we are in the position of presenting an explicit expression for the fermionic propagator using Riemann normal coordinates about $x'$. Using that $m^2 = M_{e}^2 = R(x')/12$ for fermions as well as the above expansions for $\Delta_{\textrm{VM}}(x,x')$ and $\gamma^{\mu} \nabla_{\mu}$, one finds, for the fermionic propagator
\bea
\hspace{-20mm}
S_0(x,x')  &=& \int \frac{d^{D} k}{(2\pi)^{D}} e^{- ik y} 
\left[\frac{ \gamma^{\nu} k_{\nu} }{k^2 - M_{e}^2}
+ \frac{1}{(k^2 - M_{e}^2)^{2}} 
\left( \frac{1}{2}  R_{\nu \rho } \gamma^{\nu}  k^{\rho}
-  \frac{\gamma^{\nu} k_{\nu}}{6} R \right)
\right.
\nn\\
&+& \left.  \frac{2}{3} \frac{ \gamma^{\nu} k_{\nu} k^{\sigma} k^{\rho} R_{\rho\sigma}}{(k^2 - M_{e}^2)^3} 
+ \cdots \right]
\eea
where in the above equation $\gamma^{\mu}$ is the usual gamma matrix in flat space and
$R = R_{\mu\nu} \eta^{\mu\nu}$ when considering only terms linear in the curvature for the expansion of $g^{\mu\nu}$ in Riemann normal coordinates.

\section{Local-momentum representation of the gauge propagator}
\label{B}

In this Appendix we present the local-momentum representation of the gauge propagator. For a standard discussion, see for instance Refs.~\cite{Parker:84,Toms:14,Buchbinder:1984uy}. In the present case, one has that the gauge propagator obeys Eq.~(\ref{gauge}). Following~\cite{Jack:85,Parker:86} one has, for the gauge propagator (in the Feynman gauge $\xi =1$)
\beq
G^{\mu}_{\ \nu'}(x,x') = i \int_{0}^{\infty} ds \langle x,s | x',0 \rangle^{\mu}_{\ \nu'} 
\eeq
with
\bea
\langle x,s | x',0 \rangle^{\mu}_{\ \nu'} &=& i (4\pi i s)^{-d/2} e^{i \sigma(x,x')/2s} 
\Delta_{\textrm{VM}}^{1/2}(x,x')
\bar{H}^{\mu}_{\ \nu'}(x,x'; is) e^{is R(x')/6 }
\nn\\
\bar{H}^{\mu}_{\ \nu'}(x,x'; is) &=& g^{\mu}_{\ \nu'}(x,x') + \sum_{j=1}^{\infty} (is)^{j} 
\bar{h}_{j}\,^{\mu}_{\ \nu'}(x,x')
\eea
We stress that $\bar{h}_{j}\,^{\mu}_{\ \nu'}$ is a bivector. Recall that, for a proper expansion of a bivector, such as the gauge propagator, one must form the combination $g^{\nu}_{\ \lambda'} G^{\mu\lambda'}$, which is a contravariant tensor of rank two at $x$ and a scalar at $x'$. The object $g^{\nu}_{\ \lambda'}$ is the bivector of parallel transport from $x'$ to $x$~\cite{Christensen:78}. Note that $g_{\mu\nu'}(x,x) = g_{\mu\nu}$.

Proceeding with analogous considerations as above, one obtains that
\bea
\bar{h}_{1}\,^{\mu}_{\ \nu'}(x') &=& h_{1}\,^{\mu}_{\ \nu'}(x') + B(x') g^{\mu}_{\ \nu'} 
\nn\\
\bar{h}_{1\alpha}\,^{\mu}_{\ \nu'}(x') &=& h_{1\alpha}\,^{\mu}_{\ \nu'}(x') 
\nn\\
\bar{h}_{1\alpha\beta}\,^{\mu}_{\ \nu'}(x') &=& h_{1\alpha\beta}\,^{\mu}_{\ \nu'}(x')
\nn\\
\bar{h}_{2}\,^{\mu}_{\ \nu'}(x') &=& h_{2}\,^{\mu}_{\ \nu'}(x') + \frac{1}{2} B^{2}(x') g^{\mu}_{\ \nu'}
+   B(x')h_{1}\,^{\mu}_{\ \nu'}(x') 
\eea
where $B(x') = - R(x')/6 $, $\bar{h}_{1}\,^{\mu}_{\ \nu'}(x,x') = \bar{h}_{1}\,^{\mu}_{\ \nu'}(x') 
+ \bar{h}_{1\alpha}\,^{\mu}_{\ \nu'}(x') y^{\alpha} + \bar{h}_{1\alpha\beta}\,^{\mu}_{\ \nu'}(x') y^{\alpha} y^{\beta} + {\cal O}(y^3)$. Here the coefficients $h_{j}\,^{\mu}_{\ \nu'}$ can also be found in the literature~\cite{Christensen:78}.

As above, we are interested only in terms linear in the Riemann curvature. Using Riemann normal coordinates about $x'$, one obtains
\beq
G_{\mu\nu'}(x,x') = - \Delta_{\textrm{VM}}^{1/2}(x,x') \int \frac{d^{d_{\gamma}} k}{(2\pi)^{d_{\gamma}}} 
e^{- iky} 
\bar{H}_{\mu\nu'}\left( x,x'; -\frac{\partial}{\partial M_{\gamma}^2} \right)
\frac{1}{k^2 - M_{\gamma}^2}
\eeq
where $M_{\gamma}^2 = - R(x')/6$ and we used that~\cite{Parker:86}
\beq
g_{\mu\nu'}(x,x') = \eta_{\mu\nu'} - \frac{1}{6} R_{\mu\rho\sigma\nu'}(x') y^{\rho} y^{\sigma} + \cdots .
\eeq
By using the aforementioned expansion for the Van Vleck-Morette determinant, together with previous results, one finds that
\bea
\hspace{-10mm}
G_{\mu\nu'}(x,x') &=& - 
 \int \frac{d^{d_{\gamma}} k}{(2\pi)^{d_{\gamma}}} e^{- iky} 
\left[\frac{ \eta_{\mu\nu'} }{k^2 - M_{\gamma}^2}
+ \frac{1}{(k^2 - M_{\gamma}^2)^{2}} 
\left( \frac{2}{3}  R_{\mu\nu'} - \frac{1}{6} R \eta_{\mu\nu'} \right)
\right.
\nn\\
&-& \left.  \frac{2}{3} \frac{ ( 2 R_{\mu\alpha\beta\nu'} - R_{\alpha\beta} \eta_{\mu\nu'}) 
k^{\alpha} k^{\beta}}{(k^2 - M_{\gamma}^2)^3} 
+ \cdots \right].
\eea
Recall that the gauge propagator obtained above corresponds to the one in $d_{\gamma}$ dimensions. Since here we are interested in the properties of the system in the reduced space where the fermion field is living, we integrate over the $d_{\gamma}-d_{e}$ bulk gauge degrees of freedom.


\end{document}